\documentclass[journal]{IEEEtran}
\ifCLASSINFOpdf
\else
\fi
\usepackage{epsfig,amssymb,amsmath}
\usepackage[utf8]{inputenc}
\usepackage{amsmath,graphicx}
\usepackage{amssymb,amsmath,epsfig,url,mathrsfs} \usepackage{algorithm,algorithmic}
\usepackage[usenames, dvipsnames]{color}
\usepackage[framemethod=TikZ]{mdframed}
\usepackage{xcolor}
\usepackage{colortbl}
\usepackage[colorinlistoftodos]{todonotes}
\usepackage{hyperref}
\usepackage{multirow}
\usepackage{enumitem}
\usepackage{subfigure}
\newcommand*\dif{\mathop{}\!\mathrm{d}} 
\usepackage{pgfplots}

\usepackage{graphicx}

\usepackage{tikz}
\usetikzlibrary{shapes,snakes}

\usepackage{array}
\newcolumntype{P}[1]{>{\centering\arraybackslash}m{#1}}

\renewcommand{\vec}[1]{\boldsymbol{\mathrm{#1}}}
\newcommand{\mtx}[1]{\boldsymbol{\mathrm{#1}}}

\newcommand\E{\mathbb{E}}

\renewcommand{\vec}{\boldsymbol} 

\usepackage{amsthm}

\makeatletter
\newcommand{\mathleft}{\@fleqntrue\@mathmargin0pt}
\newcommand{\mathcenter}{\@fleqnfalse}
\makeatother
 
\theoremstyle{definition}

\hypersetup{
   unicode=false,          
   pdftoolbar=true,        
   pdfmenubar=true,        
   pdffitwindow=false,     
   pdfstartview={FitH},    
   pdftitle={Tandem Assessment of Spoofing Countermeasures and Automatic Speaker Verification: Fundamentals},    
   pdfauthor={T. Kinnunen et al.},     
   pdfsubject={Subject},   
   pdfcreator={Creator},   
   pdfproducer={Producer}, 
   pdfkeywords={keyword1} {key2} {key3}, 
   pdfnewwindow=true,      
   colorlinks=true,       
   linkcolor=blue,          
   citecolor=blue,        
   filecolor=magenta,      
   urlcolor=blue           
}

\mdfdefinestyle{MyFrame}{
    linecolor=black,
    outerlinewidth=.1pt,
    roundcorner=1pt,
    innertopmargin=\baselineskip,
    innerbottommargin=\baselineskip,
    innerrightmargin=5pt,
    innerleftmargin=0pt,
    backgroundcolor=gray!10!white}

\begin{document}
%
\title{Tandem Assessment of Spoofing Countermeasures and Automatic Speaker Verification: Fundamentals}

%
%
%

\author{Tomi~Kinnunen,~\IEEEmembership{Member,~IEEE,}
H\'ector~Delgado,~\IEEEmembership{Member,~IEEE,}
Nicholas~Evans~\IEEEmembership{Member,~IEEE,}
Kong~Aik~Lee,~\IEEEmembership{Senior Member,~IEEE,}
Ville~Vestman, 
Andreas~Nautsch,~~\IEEEmembership{Member,~IEEE,}
Massimiliano~Todisco,~\IEEEmembership{Member,~IEEE,}
Xin~Wang,~\IEEEmembership{Member,~IEEE,}
Md~Sahidullah~\IEEEmembership{Member,~IEEE,}
Junichi~Yamagishi,~\IEEEmembership{Senior Member,~IEEE,}
and~Douglas~A.~Reynolds,~\IEEEmembership{Fellow,~IEEE,}
\thanks{T.\ Kinnunen and V.\ Vestman are with the School of Computing, University of Eastern Finland, Länsikatu 15, FI-80101 Joensuu, Finland. E-mail: $\{$tomi.kinnunen,ville.vestman$\}$@uef.fi.}
\thanks{H.\ Delgado is with Nuance Communications Inc., 39 Gran V\'ia, 28013 Madrid, Spain. E-mail: hector.delgado@nuance.com. Part of his contribution to this work was done during his previous role at EURECOM, France.}
\thanks{N.\ Evans, A.\ Nautsch and M.\ Todisco are with EURECOM, Campus SophiaTech, 450 Route des Chappes, 06410 Biot, France. E-mail: $\{$evans, nautsch, todisco$\}$@eurecom.fr}
\thanks{K.A.\ Lee is with NEC Corporation, 7-1, Shiba 5-chome Minato-ku, Tokyo 108-8001, Japan. E-mail: k-lee@ax.jp.nec.com}%
\thanks{X.\ Wang and J.\ Yamagishi are with National Institute of Informatics, 2-1-2 Hitotsubashi, Chiyoda-ku, Tokyo, Japan. E-mail: $\{$wangxin@nii.ac.jp,jyamagis$\}$@nii.ac.jp}%
\thanks{M. Sahidullah is with the Universit\'{e} de Lorraine, CNRS, Inria, LORIA, F-54000, Nancy, France. E-mail: md.sahidullah@inria.fr}%
\thanks{D.A.\ Reynolds is with MIT Lincoln Laboratory, Massachusetts Institute of Technology, 244 Wood Street, Lexington, MA 02421-6426, E-mail: dar@ll.mit.edu}%
\thanks{Published in IEEE/ACM-TASLP (doi: 10.1109/TASLP.2020.3009494). \copyright 2020 IEEE. Personal use of this material is permitted. Permission from IEEE must be obtained for all other uses, in any current or future media, including reprinting/republishing this material for advertising or promotional purposes, creating new
collective works, for resale or redistribution to servers or lists, or reuse of any copyrighted component of this work in other works.
}}

%
%

\markboth{Published in IEEE/ACM Transactions on Audio, Speech, and Language Processing (10.1109/TASLP.2020.3009494)}%
{Kinnunen \MakeLowercase{\textit{et al.}}: Tandem Assessment of Spoofing Countermeasures}

%



\maketitle

\begin{abstract}
Recent years have seen growing efforts to develop spoofing countermeasures (CMs) to protect automatic speaker verification (ASV) systems from being deceived by manipulated or artificial inputs. The reliability of spoofing CMs is typically gauged using the equal error rate (EER) metric. The primitive EER fails to reflect application requirements and the impact of spoofing and CMs upon ASV and its use as a primary metric in traditional ASV research has long been abandoned in favour of risk-based approaches to assessment. This paper presents several new extensions to the tandem detection cost function (t-DCF), a recent risk-based approach to assess the reliability of spoofing CMs deployed in tandem with an ASV system. Extensions include a simplified version of the t-DCF with fewer parameters, an analysis of a special case for a fixed ASV system, simulations which give original insights into its interpretation and new analyses using the ASVspoof 2019 database. It is hoped that adoption of the t-DCF for the CM assessment will help to foster closer collaboration between the anti-spoofing and ASV research communities.
\end{abstract}


%
\IEEEpeerreviewmaketitle

%
%
%
%

\section{Introduction}

\IEEEPARstart{B}iNARY classifiers (or detectors) are prone to two different types of errors, \emph{misses} and \emph{false alarms}. For biometric recognition systems such as \emph{automatic speaker verification} (ASV) used for authentication, miss and false alarm rates are proxies for user convenience and security, respectively. User convenience and security are competing requirements. The compromise between them will depend upon the application; whereas an online banking application might call for high security, user convenience might be key to a successful smarthome application. The approach to assessment must hence reflect application considerations. One such approach to assessment, used for the standard ASV evaluation benchmarks run by the \emph{National Institute of Standards and Technology} (NIST) in the US since 1996 \cite{Greenberg2020_twenty_years}, is the \emph{detection cost function}~(DCF) \cite{Doddington2000-NIST-overview}.   

\begin{figure}[!t]
	\centering
  \includegraphics[width=0.50\textwidth]{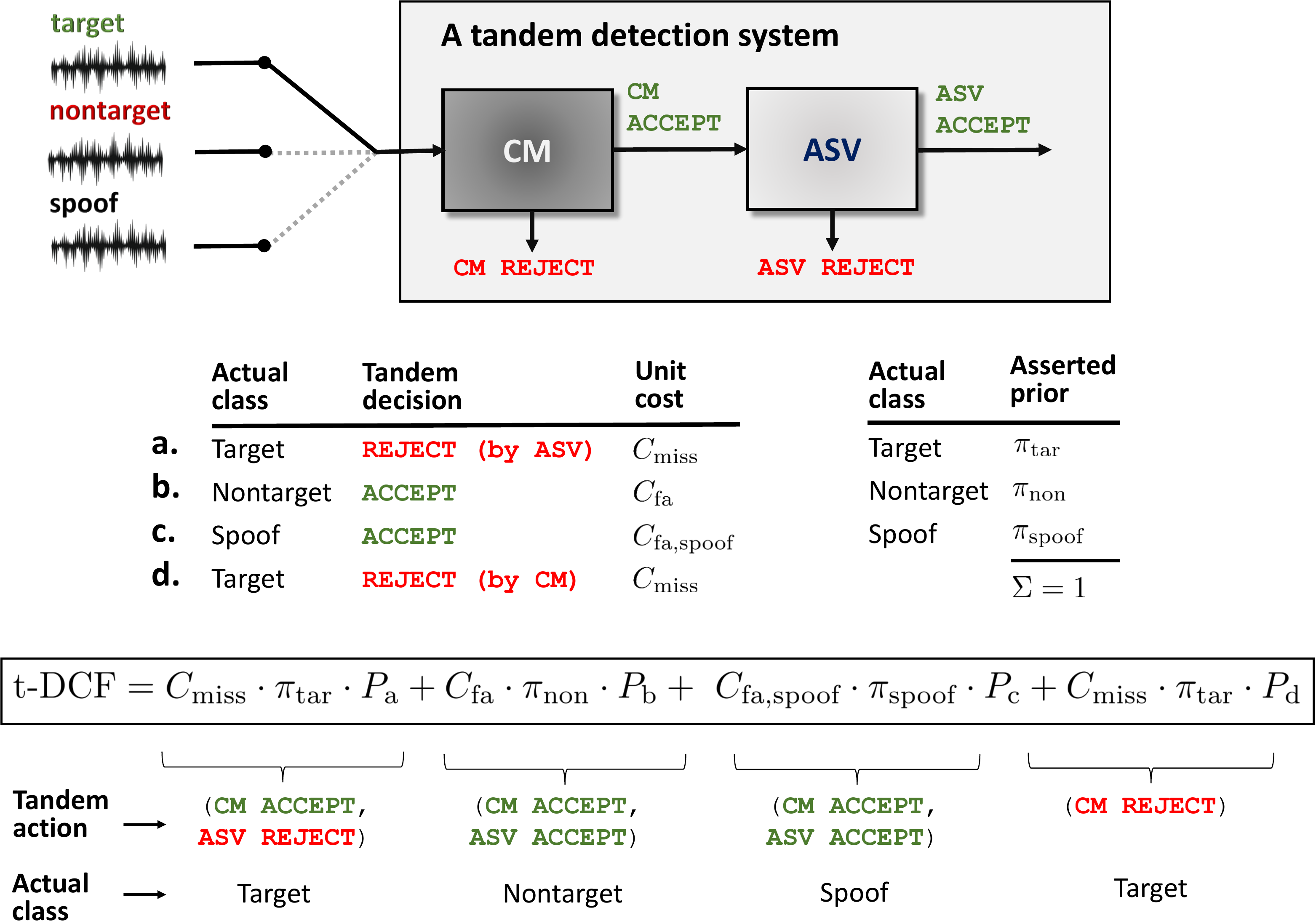}
	\caption{A tandem system consisting of \emph{automatic speaker verification} (ASV) and \emph{spoofing countermeasure} (CM) modules is evaluated using three types of trials: targets, nontargets and spoofing attacks. \emph{Tandem detection cost function} (t-DCF) is the average cost of erroneous decisions by the tandem system.}
	\label{fig:tDCF-idea-in-one-picture}
\end{figure}

The DCF reflects the cost of decisions in a \emph{Bayes risk} sense~\cite{DudaHartStork01,Jaynes03} and was designed for the assessment of ASV systems with a protocol involving a mix of \emph{target} and \emph{non-target} trials. The latter are casual impostors who make no effort to fool the ASV system. This paper concerns the assessment of ASV systems in the face of fake, falsified or spoofed inputs, also referred to as \emph{presentation attacks}~\cite{isopad}.  These are specially crafted inputs that are used by a fraudster to deceive an ASV system and hence to provoke false alarms.  Just like all other biometrics systems, ASV systems can be vulnerable to spoofing~\cite{Interspeech2013SS}.  The usual means to defend against such attacks involves the coupling of ASV systems with spoofing \emph{countermeasures}~(CMs) \cite{Satoh2001-HMM-imposture}, namely sub-systems designed to distinguish not between target and non-target trials, but between genuine, human or bona fide speech and artificially generated or manipulated inputs.

Since 2013, research in anti-spoofing for ASV has been spearheaded through the community-led ASVspoof initiative. It has produced three, well-supported competitive challenges. The first two editions held in 2015 and 2017 used an \emph{equal error rate}~(EER) metric to assess the performance of \emph{isolated} spoofing CMs.  Just like ASV systems, CMs are binary classifiers and; they make two types of errors that also have different consequences, depending on the application.  The approach to assessment should hence reflect differences in the cost of each type of error.  Furthermore, CMs are always used in combination with ASV.  Previous work has shown the potential to combine the action of ASV and CM systems in the form of a single, \emph{integrated} system~\cite{Sizov2015-tifs}, by the back-end \emph{fusion} of independently trained ASV and CM systems~\cite{Sahid2016-integrated,Todisco2018-integrated}, or via the \emph{tandem} detection framework illustrated in Fig.~\ref{fig:tDCF-idea-in-one-picture}.  Each approach shares the common goal of protecting ASV systems from being deceived by spoofed inputs. Since CMs are never used alone, the use of EER to assess the performance of spoofing CMs is hence questionable. We need better approches to assess CM performance, specifically approaches which reflect application requirements.

In order to preserve trust, ASV systems used for authentication in sensitive applications should have the capacity to defend against spoofing.  In this sense, research in ASV and anti-spoofing is inextricably intertwined.  Unfortunately, though, the two communities are today somewhat disjointed.  One explanation for this situation might stem from the disparity between the metrics used in each field.  It is difficult to argue, though, that the two communities should not work together, for they share the same goal to develop ever-more reliable ASV technology.  In trying to foster closer collaboration, the work presented in this paper explores how the infrastructure and metrics developed for the assessment of ASV can be adopted for the joint assessment of ASV and spoofing CMs.  It presents the \emph{tandem detection cost function} (t-DCF) approach to assess the performance of tandem CM and ASV systems of the form illustrated in Fig.~\ref{fig:tDCF-idea-in-one-picture}.

A preliminary version of this work was presented in \cite{Kinnunen2018-tDCF}. The current work extends it in a number of respects. First, we revisit the t-DCF considering a reduced set of five evaluation parameters (rather than seven). Second, in contrast to the \emph{unconstrained} t-DCF formulation for which both ASV and CM system thresholds can be varied, 
we present the \emph{ASV-constrained} t-DCF which shows how a CM should be optimised for a given, fixed ASV system with known miss, false alarm and \emph{spoof false alarm} (SFAR) rates. Third, we present a numerical simulation of the t-DCF which sheds light upon its behaviour and helps to interpret results.  Fourth, whereas \cite{Kinnunen2018-tDCF} presented results for ASVspoof 2015 and 2017 datasets, the current work presents new results and analysis for the most recent ASVspoof 2019 datasets. Finally, the paper contains several new clarifications and further examples not presented in~\cite{Kinnunen2018-tDCF}. The material is intended as a self-contained and accessible introduction for both experts and non-experts. While the paper relates to ASV, it should be of interest to the broader biometrics community, where work in anti-spoofing evaluation remains characterised by the use of \emph{ad-hoc} bases rather than an application-targeted metric.

\begin{table*}[!t]
\caption{The ingredients of t-DCF. The system illustrated in Fig. \ref{fig:tDCF-idea-in-one-picture} can face three types of trials (first column): \emph{target}, \emph{nontarget} and \emph{spoof}, each with some asserted \textbf{prior} (second column). The system either rejects or accepts the trial by taking a \textbf{tandem action} (third column), with nonnegative \textbf{detection costs} (fourth column) assigned to each error (0: no loss). \textbf{Detection error rate} of the tandem system (last column) is obtained by treating the CM and ASV system decisions independent. The t-DCF metric is obtained by multiplying the prior (\emph{how often we expect to observe this class?}), cost (\emph{if this error happens, how much does it cost?}) and the error rate (\emph{how many errors were actually observed?}) of each row and summing up the terms.}\label{tab:tDCF-ingredients}
\begin{center}
    \begin{tabular}{cc|lc|c}
        Actual class & Class prior & \textbf{Tandem action} = (CM action, ASV action)& Detection Cost & Detection error rate\\
        \hline\hline
        \multirow{3}{*}{Target} & \multirow{3}{*}{$\pi_\text{tar}$} & $\alpha_1=\textbf{\texttt{REJECT}}=(\texttt{CM ACCEPT},\texttt{ASV REJECT})$ & $C_\text{miss}$ & $(1-P_\text{miss}^\text{cm}(\tau_\text{cm}) )\times P_\text{miss}^\text{asv}(\tau_\text{asv})$\\
                &  & $\alpha_2=\textbf{\texttt{ACCEPT}}=(\texttt{CM ACCEPT},\texttt{ASV ACCEPT})$ & 0 & --\\
                &  & $\alpha_3=\textbf{\texttt{REJECT}}=(\texttt{CM REJECT})$ & $C_\text{miss}$ & $P_\text{miss}^\text{cm}(\tau_\text{cm})$\\
        \hline
        \multirow{3}{*}{Nontarget} & \multirow{3}{*}{$\pi_\text{non}$} & $\alpha_1=\texttt{\textbf{REJECT}}=(\texttt{CM ACCEPT},\texttt{ASV REJECT})$ & 0 & --\\
                &  & $\alpha_2=\textbf{\texttt{ACCEPT}}=(\texttt{CM ACCEPT},\texttt{ASV ACCEPT})$ & $C_\text{fa}$ & $(1-P_\text{miss}^\text{cm}(\tau_\text{cm}))\times P_\text{fa}^\text{asv}(\tau_\text{asv})$\\
                &  & $\alpha_3=\textbf{\texttt{REJECT}}=(\texttt{CM REJECT})$ & 0 & --\\
        \hline
        \multirow{3}{*}{Spoof} & \multirow{3}{*}{$\pi_\text{spoof}$} & $\alpha_1=\textbf{\texttt{REJECT}}=(\texttt{CM ACCEPT},\texttt{ASV REJECT})$ & 0 & --\\
                &  & $\alpha_2=\textbf{\texttt{ACCEPT}}=(\texttt{CM ACCEPT},\texttt{ASV ACCEPT})$ & $C_\text{fa,spoof}$ & $P_\text{fa}^\text{cm}(\tau_\text{cm})\times P_\text{fa,spoof}^\text{asv}(\tau_\text{asv})$\\
                &  & $\alpha_3=\textbf{\texttt{REJECT}}=(\texttt{CM REJECT})$ & 0 & --\\
        \hline
    \end{tabular}
\end{center}
\end{table*}

\section{The two systems and their tandem combination}\label{sec:the-two-systems}

Both \emph{automatic speaker verification} (ASV) and \emph{spoofing countermeasure} (CM) systems are binary classifiers. An ASV trial consists of an enrollment-test utterance pair $(\mathcal{X}_\text{e},\mathcal{X}_\text{t})$ where $\mathcal{X}_\text{e}$ is collected at the enrollment stage and $\mathcal{X}_\text{t}$ at the verification stage. A pair $(\mathcal{X}_\text{e},\mathcal{X}_\text{t})$ 
with matched speaker identities is known as a \emph{target} trial, otherwise as a \emph{non-target} trial. The ASV system propositions (hypotheses) are hence given by:
\begin{equation}\nonumber\label{eq:asv-hypotheses}
    \left\{\begin{aligned}
    H_0^\text{asv} & \text{\,(\textbf{target} hypothesis)}       :\,\,\, \mathsf{id}(\mathcal{X}_\text{e}) =\mathsf{id}(\mathcal{X}_\text{t})\\
    H_1^\text{asv} & \text{\,(\textbf{nontarget} hypothesis)}       :\,\,\, \mathsf{id}(\mathcal{X}_\text{e}) \neq\mathsf{id}(\mathcal{X}_\text{t}),\\
    \end{aligned}
    \right.
\end{equation}
where $\mathsf{id}(\mathcal{X}) \in \mathbb{N}=\{1,2,\dots\}$ is the unique speaker identity (a categorical variable) of utterance $\mathcal{X}$. The ASV system can encounter spoofed trials too. It is because the ASV system is assumed to have limited (or no) capacity to reject spoofs that dedicated CMs are needed. 

The CM operates only upon the test utterance $\mathcal{X}_\text{t}$ and aims to verify its authenticity. If it corresponds to genuine speech produced by a human speaker, then the test upon $\mathcal{X}_\text{t}$ performed by the CM is referred to as a \emph{bona fide} trial. If it corresponds to non-genuine, manipulated or synthesized speech, then it is referred to as a \emph{spoof} trial.  The CM propositions are hence given by:
\begin{equation}\nonumber
    \left\{\begin{aligned}
    H_0^\text{cm} & \text{\,(\textbf{bona fide} hypothesis)}     : \text{$\mathcal{X}_\text{t}$ is bona fide speech}\\
    H_1^\text{cm} & \text{\,(\textbf{spoof} hypothesis)}         : \text{$\mathcal{X}_\text{t}$ is spoofing attack.}\\
    \end{aligned}
    \right.
\end{equation}

The CM system is designed to distinguish bona fide from spoof trials. In the same way that the ASV system has limited capacity to reject spoofing attacks, the CM is assumed to have limited capacity to distinguish target from nontarget trials; both are bona fide. The ASV and CM systems play complementary roles and both are needed to ensure spoofing-robust ASV.

We define the \textbf{tandem system} as a cascade of CM and ASV systems, as illustrated in Fig.~\ref{fig:tDCF-idea-in-one-picture}. The CM acts as a \emph{gate} which aims to prevent spoofing attacks from reaching the ASV system. Conventional ASV systems can also be regarded as tandem systems with a dummy `accept all' CM~\cite{Kinnunen2018-tDCF}.  Accordingly, the work presented applies also to the analysis of conventional ASV systems. \emph{Internally} the tandem system consists of two subsystems that act together (and whose errors combine). To \emph{end users} the tandem system acts as a single ASV system that either accepts or rejects their identity claim; the tandem system should therefore be viewed as a spoofing-robust ASV system with a CM `under the hood'. The tandem system can encounter three different types of trials: (i)~\emph{target}, (ii)~\emph{nontarget} and (iii)~\emph{spoof}. It should accept only the target trials. Both nontarget and spoof trials should be rejected.

\begin{figure*}
\subfigure[Automatic speaker verification (ASV).]{\includegraphics[width=.48\textwidth]{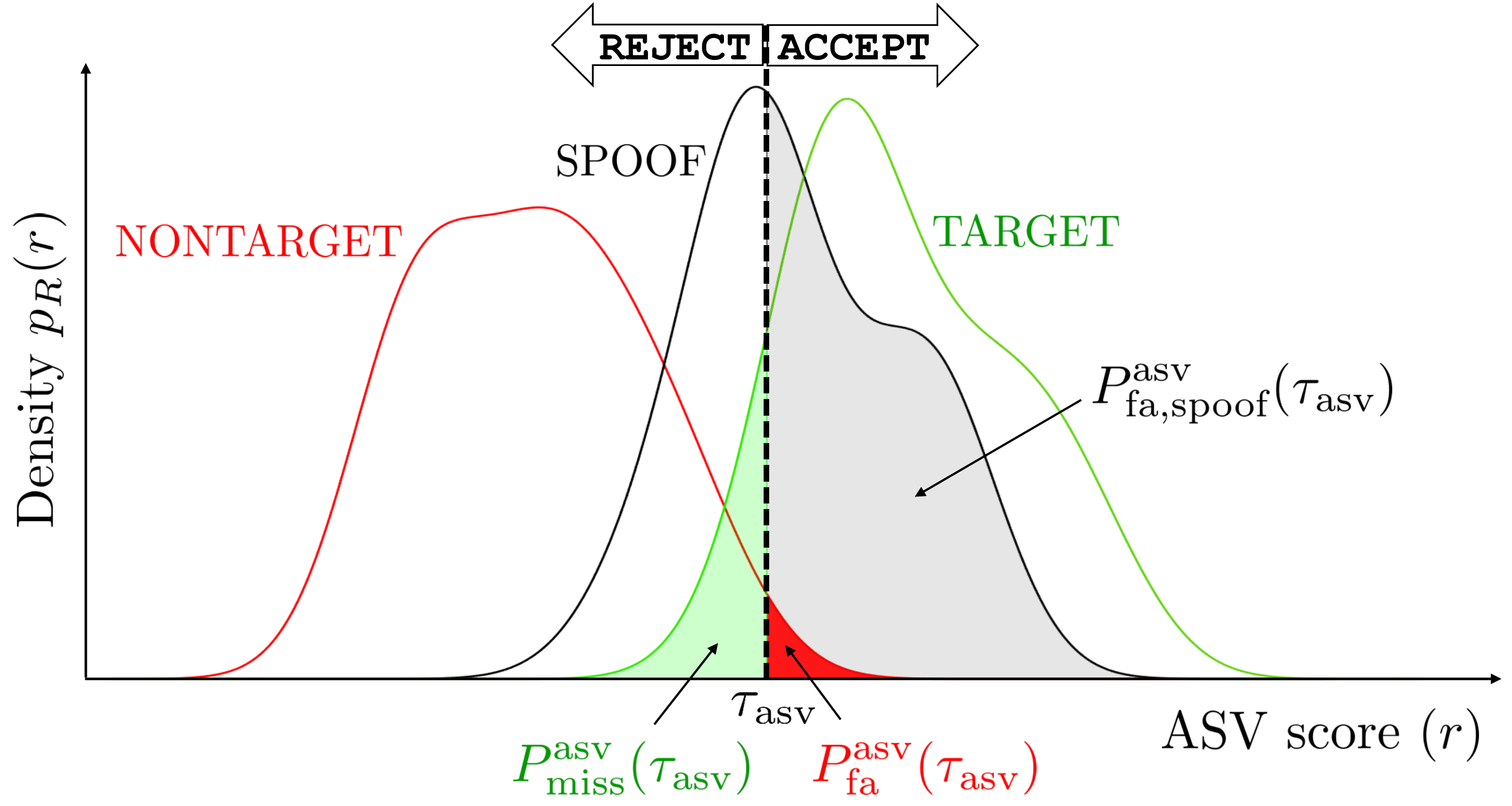}}
\subfigure[Spoofing countermeasure (CM).]{\includegraphics[width=.48\textwidth]{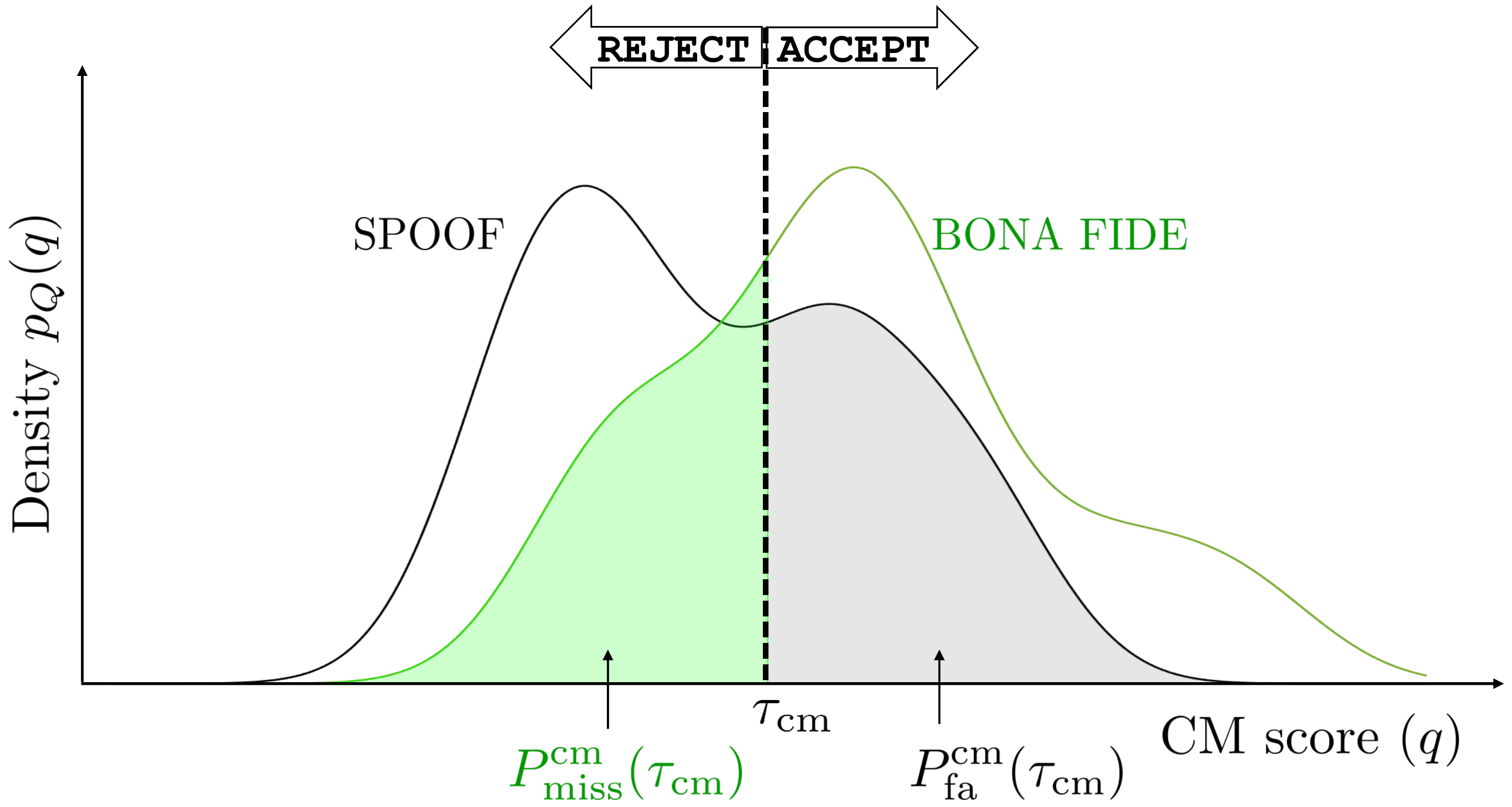}}
\caption{The ASV and CM score density functions $p_R(r)$ and $p_Q(q)$, respectively. Fixing the detection thresholds of ASV $(\tau_\text{asv})$ and CM $(\tau_\text{cm})$ yields five different types of error rates: miss and false alarm rates of each system and spoof false alarm rate of the ASV system. The error rates of ASV and CM are multiplied to form tandem error rates (see Table \ref{tab:tDCF-ingredients}).}\label{fig:score-schematic}
\end{figure*}

\section{Tandem detection cost function}

The \emph{tandem detection cost function} (t-DCF)~\cite{Kinnunen2018-tDCF} metric reflects the performance of a combined ASV and CM system for an assumed operating environment (application). Unlike the EER, the t-DCF is a \emph{parametric} function which requires the specification of application parameters in advance. 

\subsection{Detection Costs and Priors (the Application)}

Consider a hypothetical `banking' scenario in which customer authentication is controlled using voice biometrics. Access should be restricted to target users (account owners), while nontarget (zero-effort impostor) and spoofed (dedicated impostor) access attempts should always be denied; the system should be \emph{secure}. Access by customers should always be granted and never denied; they should not be \emph{inconvenienced}. The competing requirements for security and convenience cannot both be satisfied, leading to the potential for detection errors. To each error is associated a monetary loss (\emph{e.g.}\ loss of funds to fraud, or loss of customers to inconvenience). 

In seeking to minimise its costs, the bank will assign a higher penalty to the more costly errors. This is formalized through the specification of \emph{detection costs}, $C(\alpha|\theta) \geq 0$, interpreted as the penalty of taking \emph{action} $\alpha$ (making a 
decision for a given test trial) when the actual class is $\theta$ \cite{DudaHartStork01,Jaynes03}. Correct decisions are assigned a cost of 0 while erroneous decisions are assigned a positive numerical value, which signifies the monetary loss to the bank incurred as a result of each type of detection error.

By denoting the class variable by $\theta \in \Theta \equiv \{\theta_\text{tar},\theta_\text{non},\theta_\text{spoof}\}$ and the action by $\alpha \in \mathcal{A} \equiv \{\texttt{ACCEPT},\texttt{REJECT}\}$, we
define the following three detection costs:
\begin{itemize}
	\item $C_\text{miss}\equiv C(\texttt{REJECT}|\theta_\text{tar})$ -- cost of rejecting a target trial;
    \item $C_\text{fa}\equiv C(\texttt{ACCEPT}|\theta_\text{non})$ -- cost of accepting a nontarget trial;
    \item $C_\text{fa,spoof}\equiv C(\texttt{ACCEPT}|\theta_\text{spoof})$ -- cost of accepting a spoofed trial,
\end{itemize}
where the first two correspond to the familiar notations used in NIST \emph{speaker recognition evaluation} (SRE) campaigns~\cite{Greenberg2020_twenty_years,Doddington2000-NIST-overview}. The third cost is specific to the new class of spoofing attacks. It is stressed that actions are \emph{those of the tandem system}. This is different to~\cite{Kinnunen2018-tDCF} where costs are specified \emph{per subsystem}, but  error rates are those of the tandem system. 

The first type of tandem error occurs when either the CM or ASV system rejects a target; the second case occurs when both systems accept a nontarget; the last case occurs when both systems accept a spoofing attack. As displayed in Table \ref{tab:tDCF-ingredients}, these four cases cover all the possible errors. The remaining five cases lead to correct tandem decisions and are therefore assigned zero cost. Note the curious cases of nontargets being rejected by the CM system (6th row), and spoofing attacks rejected by the ASV system (7th row). These trials are rejected by the `wrong' subsystem but neither incurs loss as the tandem action is correct.

In addition to assigning detection costs for each type of error, one must also take into account the relative occurences of the three classes (target, nontarget, spoof).
A relatively expensive error that occurs only infrequently may cost less than a relatively inexpensive error that occurs more frequently. For instance, if the bank expects 99\% of authentication requests to originate from bona fide account holders, then the expected monetary loss incurred from bona fide customers (target trials) being denied access to their accounts may override the costs incurred from access being granted erroneously to fraudsters (zero-effort or spoofed trials), who account for only 1\% of authentication requests. The assumed commonality of each trial class is encoded in their \emph{prior probabilities}:  
\begin{itemize}
	\item $\pi_\text{tar} \equiv P_\Theta(\theta_\text{tar})$ -- prior probability of target;
    \item $\pi_\text{non} \equiv P_\Theta(\theta_\text{non})$ -- prior probability of nontarget;
    \item $\pi_\text{spoof} \equiv P_\Theta(\theta_\text{spoof})$ -- prior probability of spoofing attack,
\end{itemize}
where $P_\Theta(\theta)$ is a shorthand for $P_\Theta(\Theta=\theta)$, $\Theta$ and $\theta$ being a random variable and its realization, respectively. 

The priors are nonnegative and sum to unity (therefore, fixing any two priors automatically defines the third). The prior is \emph{subjective} --- it asserts the belief of the relative frequency of each trial class during the operation of a spoofing-robust ASV (with the actual, empirical class frequencies remaining unknown). The priors of the cost function do not have to (and typically do not) correspond to the empirical trial frequencies in training or evaluation corpora. The costs and priors are set in advance and they remain fixed within a given evaluation or application setting; they might be very different for, \emph{e.g.}, banking, forensics or surveillance applications. The set of evaluation metric parameters are hence given by $\vec{\Psi}_\text{t-DCF}\equiv (\pi_\text{tar},\pi_\text{spoof},C_\text{miss},C_\text{fa},C_\text{fa,spoof})$, where the nontarget prior is omitted and obtained from $\pi_\text{non}=1-\pi_\text{tar}-\pi_\text{spoof}$.  

\subsection{Detection Error Rates of ASV and CM}

Let $r=\mathsf{ASV}(\mathcal{X}_\text{e},\mathcal{X}_\text{t})$ and $q=\mathsf{CM}(\mathcal{X}_\text{t})$ denote ASV and CM scores\footnote{Here $\mathsf{ASV}(\cdot,\cdot)$ and $\mathsf{CM}(\cdot)$ denote the two system operations as `black-boxes' that gives us the detection scores. Usually $r$ and $q$ are the \emph{logarithm of the likelihood ratio} (LLR) of the respective null and alternative hypothesis likelihoods obtained from a parametric model (though the proposed metric does not require a LLR interpretation of scores).}, treated here as realizations of random variables that admit continuous probability density functions $p_R(r)$ and $p_Q(q)$. That is, $p_R(r)\geq 0, \int_{-\infty}^\infty p_R(r)\dif r=1$ (similarly for $p_Q(q)$). 
The testing of the ASV system with all three types of trials leads to detection scores 
drawn from the respective class-conditional distributions:
    \begin{equation}\label{eq:asv-probability-densities}
        \begin{aligned}
            p_R(r|\theta_{\text{tar}}) & \text{\,\,\,\,\,{\small ASV target score distribution;}}\\ 
            p_R(r|\theta_{\text{non}}) & \text{\,\,\,\,\,{\small ASV  nontarget score distribution;}}\\
         p_R(r|\theta_{\text{spoof}}) & \text{\,\,\,\,\,{\small ASV spoof score distribution.}}\\
        \end{aligned}
    \end{equation}
Likewise, testing of the CM with trials of bona fide and spoof classes leads to detection scores 
drawn from the conditional distributions
    \begin{equation}\label{eq:cm-probability-densities}
     \begin{aligned}
           p_Q(q|\theta_{\text{bona}}) & \text{\,\,\,\,\,{\small CM bona fide score distribution;}}\\ 
            p_Q(q|\theta_{\text{spoof}}) & \text{\,\,\,\,\,{\small CM spoof score distribution,}}
        \end{aligned}
    \end{equation}
where we introduced $\theta_{\text{bona}}$ as a realization of a new random variable $\Theta_{\text{bona}} \equiv \{\theta_\text{tar}\} \cup \{\theta_\text{non}\} \subset \Theta$, a container of any \emph{non-spoof} trials (whether target or nontarget). To obtain the distribution of the bona fide scores, consider the joint distribution of $Q$ and $\Theta_\text{bona}$, $P(q,\theta_\text{bona})=P(\theta_\text{bona})p(q|\theta_\text{bona})$, obtained 
using the \emph{product rule} \cite[Eq. (1.11)]{Bishop2006} (subscripts omitted for brevity). By treating $\Theta_\text{bona}$ as a \emph{latent variable}, the bona fide score distribution is obtained by marginalizing the class variable out, using the \emph{sum rule} \cite[Eq. (1.10)]{Bishop2006}:
    \begin{equation}\label{eq:bona-fide-mixture-distribution}
        \begin{aligned}
        p_Q(q|\theta_\text{bona}) & = \sum_{\theta \in \Theta_\text{bona}} P(q,\theta)=\sum_{\theta \in \Theta_\text{bona}}P_{\Theta_{\text{bona}}}(\theta)p_Q(q|\theta)\\
        & = P_{\Theta_\text{bona}}(\theta_\text{tar})p_Q(q|\theta_\text{tar}) + P_{\Theta_\text{bona}}(\theta_\text{non})p_Q(q|\theta_\text{non})\\
        & = \tilde{\pi}_\text{tar}p_Q(q|\theta_\text{tar}) + (1-\tilde{\pi}_\text{tar})p_Q(q|\theta_\text{non}),
        \end{aligned}
    \end{equation}
which is a two-component \emph{mixture distribution} where $\tilde{\pi}_\text{tar}\equiv P_{\Theta_\text{bona}}(\theta_\text{tar})$ is the relative proportion of target trials within the bona fide class (known by the corpus designer, but not necessarily by the evaluee)\footnote{Again, this is not necessarily the same as the proportion dictated by the t-DCF priors, \emph{i.e.} $\pi_\text{tar}/(\pi_\text{tar}+\pi_\text{non})$.}. While \eqref{eq:bona-fide-mixture-distribution} represents the general form of the bona fide score distribution, the target and nontarget score distributions are typically highly overlapped, as speaker-independent CMs are usually not designed to discriminate between them. In the limiting case when the two become indistinguishable, \emph{i.e.} $p_Q(q|\theta_\text{tar})=p_Q(q|\theta_\text{non})$, \eqref{eq:bona-fide-mixture-distribution} collapses either to the target or nontarget distribution, making \emph{bona fide} an unnecessary relabeling of the original class labels.

As illustrated in Fig. \ref{fig:score-schematic}, ASV and CM systems convert detection scores $r$ and $q$ into hard binary decisions by comparing their values to preset detection thresholds denoted by $\tau_\text{asv}$ and $\tau_\text{cm}$, respectively. The ASV system takes an \texttt{ACCEPT} action if and only if $r > \tau_\text{asv}$ (otherwise \texttt{REJECT}) whereas the CM takes an \texttt{ACCEPT} action if and only if $q > \tau_\text{cm}$ (otherwise \texttt{REJECT}). Score distributions in \eqref{eq:asv-probability-densities} and \eqref{eq:cm-probability-densities} combined with fixed decision thresholds $\tau_\text{asv}$ and $\tau_\text{cm}$ yield a set of five error rates illustrated in Fig. \ref{fig:score-schematic}:
    \begin{equation}\label{eq:asv-cm-detection-errors}
        \begin{aligned}
            P_\text{miss}^\text{asv}(\tau_\text{asv}) =\int_{-\infty}^{\tau_\text{asv}}p_R(r|\theta_{\text{tar}})\dif r
            & \text{\,\,\,\,\,{\small ASV miss rate;}}\\ 
            P_\text{fa}^\text{asv}(\tau_\text{asv})=\int_{\tau_\text{asv}}^\infty p_R(r|\theta_{\text{non}})\dif r
            & \text{\,\,\,\,\,{\small ASV false alarm rate;}}\\
            P_\text{fa,spoof}^\text{asv}(\tau_\text{asv}) \equiv \int_{\tau_\text{asv}}^\infty p_R(r|\theta_{\text{spoof}})\dif r
            & \text{\,\,\,\,\,{\small ASV spoof false alarm rate;}}\\
            P_\text{miss}^\text{cm}(\tau_\text{cm}) =\int_{-\infty}^{\tau_\text{cm}}p_Q(q|\theta_{\text{bona}})\dif q
            & \text{\,\,\,\,\,{\small CM miss rate;}}\\
            P_\text{fa}^\text{cm}(\tau_\text{cm})=\int_{\tau_\text{cm}}^\infty p_Q(q|\theta_{\text{spoof}})\dif q
            & \text{\,\,\,\,\,{\small CM false alarm rate.}}\\
        \end{aligned}
    \end{equation}
The first and last two components are the usual miss and false alarm rates of each system, while the third \textbf{defines} the \emph{spoof false alarm rate} (SFAR) of the ASV system as the proportion of spoofing attacks accepted by the ASV system\footnote{Here spoofs are treated as the negative class (similar to nontargets) but this convention is actually arbitrary; as we did in \cite{Kinnunen2018-tDCF}, they could also be defined as a positive class (similar to targets) leading to an equivalent definition of SFAR as the probability of `ASV does NOT miss a spoof'.}. 

Since the probability density functions are unknown, practical computations must be performed using empirical detection scores $r_i \sim p_R(r)$ and $q_j \sim p_Q(q)$, where $i$ and $j$ index ASV and CM trials respectively. Assuming that the scores produced by each system are independent and identically distributed (i.i.d.) draws from the respective distribution, integrals in (\ref{eq:asv-cm-detection-errors}) may be replaced by summations, illustrated here for the ASV miss rate:
    \begin{equation}
        P_\text{miss}^\text{asv}(\tau_\text{asv})=\E_{r \sim p_R(r|\theta_{\text{tar}})}\left[\mathbb{I}(r < \tau_\text{asv}) \right]\approx \frac{1}{N_\text{tar}}\sum_{i \in \Lambda_\text{tar}}\mathbb{I}(r_i < \tau_\text{asv}),
    \end{equation}
where $\E_{z \sim p_Z(z)}\left[f(z)\right]$ denotes the expected value of function $f(z)$ w.r.t. distribution $p_Z(z)$, $\mathbb{I}(\cdot)$ is an indicator function that equals 1 for a true proposition and 0 otherwise, $\Lambda_\text{tar}$ indices the target trials, and $N_\text{tar}=|\Lambda_\text{tar}|$ denotes the total number of target trials. The other four error rates are computed similarly. The CM miss rate is computed by pooling CM target and nontarget scores, as both score sets are viewed as samples from the same bona fide score distribution \eqref{eq:bona-fide-mixture-distribution}. 

\subsection{The t-DCF and Its Relation to the NIST DCF}\label{sec:tDCF}

The t-DCF is a measure of the expected (average) cost of all errors made by a tandem system. The following treatment assumes the cascaded setup illustrated in Fig.~\ref{fig:tDCF-idea-in-one-picture}. 
There are three possible \emph{actions} (tandem decisions), $\alpha_i \in \mathcal{A}=\{\alpha_1,\alpha_2,\alpha_3\}$:
	\begin{equation}\label{eq:possible-actions}
	\begin{aligned}
    \alpha_1 & = (\texttt{CM ACCEPT}\, ,\texttt{ ASV REJECT})\\
    \alpha_2 & = (\texttt{CM ACCEPT}\, ,\texttt{ ASV ACCEPT})\\
    \alpha_3 & = (\texttt{CM REJECT}),
    \end{aligned}
    \end{equation}
where the last case indicates rejection of a trial at the CM stage. In this case, the ASV action is null or undefined (referred to in~\cite{Kinnunen2018-tDCF} as a dummy \texttt{SLEEP} action). As Table~\ref{tab:tDCF-ingredients} indicates, five of the nine possible (ground truth, action) 
combinations lead to correct tandem decisions. The remaining four terms all constitute errors:
\begin{enumerate}[itemsep=0ex,partopsep=1.2ex,parsep=1ex,label=(\alph*)]
    \item $(\alpha_1,\theta_\text{tar})$: the CM does not miss a bona fide trial, but the ASV system misses a target.
    \item $(\alpha_2,\theta_\text{non})$: the CM does not miss a bona fide trial, but the ASV system falsely accepts a nontarget.
    \item $(\alpha_2,\theta_\text{spoof})$: both CM and ASV systems falsely accept a spoof. 
    \item $(\alpha_3,\theta_\text{tar})$: the CM misses a target, which is never processed by the ASV system. 
\end{enumerate}
If CM and ASV error probabilities are assumed to be independent (discussed further in Subsection \ref{subsec:asv-cm-independence}), then the probability of each of these four outcomes may be expressed in terms of the error rates in \eqref{eq:asv-cm-detection-errors} as follows:
\begin{equation}\label{eq:tandem-error-probabilities}
\begin{aligned}
    P_\text{a}(\tau_\text{cm},\tau_\text{asv}) & = (1 - P^\text{cm}_\text{miss}(\tau_\text{cm})) \times P^\text{asv}_\text{miss}(\tau_\text{asv})\\
    P_\text{b}(\tau_\text{cm},\tau_\text{asv}) & = (1 - P^\text{cm}_\text{miss}(\tau_\text{cm})) \times P^\text{asv}_\text{fa}(\tau_\text{asv})\\
    P_\text{c}(\tau_\text{cm},\tau_\text{asv}) & = P^\text{cm}_\text{fa}(\tau_\text{cm}) \times  P^\text{asv}_\text{fa,spoof}(\tau_\text{asv})\\
    P_\text{d}(\tau_\text{cm},\tau_\text{asv}) & = P^\text{cm}_\text{miss}(\tau_\text{cm}).
\end{aligned}
\end{equation}
Finally, the t-DCF is obtained by multiplying the class prior, cost and error terms in each non-zero row (rows with nonzero entry in the 4th column) of Table~\ref{tab:tDCF-ingredients} and summing up the resulting four terms:
\begin{mdframed}[style=MyFrame]
\center{\textbf{Unconstrained t-DCF}}
\begin{equation}\label{eq:tDCF-unconstrained}
	\begin{aligned}
    & \text{t-DCF}(\tau_\text{cm},\tau_\text{asv}) = \\
    & C_\text{miss}\cdot \pi_\text{tar} \cdot \left[P_\text{a}(\tau_\text{cm},\tau_\text{asv})+P_\text{d}(\tau_\text{cm},\tau_\text{asv})\right]\\
         & + C_\text{fa} \cdot \pi_\text{non}\cdot P_\text{b}(\tau_\text{cm},\tau_\text{asv})\\
         & +{\ } C_\text{fa,spoof} \cdot \pi_\text{spoof} \cdot P_\text{c}(\tau_\text{cm},\tau_\text{asv}),
	\end{aligned}
\end{equation}
\end{mdframed}
where the three lines correspond to target, nontarget and spoof related errors, respectively. Eq.~(\ref{eq:tDCF-unconstrained}) is referred to as the `unconstrained' t-DCF to distinguish this formulation from a special case discussed shortly. 

It is readily seen that the NIST DCF \cite{Doddington2000-NIST-overview}, 
	\begin{equation}\label{eq:NIST-DCF}
    	\text{DCF}_\text{NIST}(\tau_\text{asv}) \equiv C_\text{miss}\pi_\text{tar}P_\text{miss}^\text{asv}(\tau_\text{asv}) + C_\text{fa}(1 - \pi_\text{tar})P_\text{fa}^\text{asv}(\tau_\text{asv}),
    \end{equation}
is a special case of the t-DCF by assuming (a) the use of a dummy `accept all' CM ($\tau_\text{cm}=-\infty \Rightarrow P_\text{miss}^\text{cm}(\tau_\text{cm})=0,P_\text{fa}^\text{cm}(\tau_\text{cm})=1$) \textbf{and} (b) there are no spoofing attacks ($\pi_\text{spoof}=0$). In this sense, the NIST DCF could be considered as an \emph{optimistic} t-DCF. Even if the NIST DCF has been instrumental to developments in the ASV field, it may not be well suited to operational environments where there is potential for spoofing attacks. 

\subsection{ASV-Constrained t-DCF}

In the unconstrained t-DCF \eqref{eq:tDCF-unconstrained} both CM and ASV are adjustable, which can make evaluation of the tandem system cumbersome. Here we explore the t-DCF from the perspective of a CM developer who cannot interact with the ASV system (or has no capacity to develop one). The ASVspoof 2019 challenge is representative of such a scenario. The evaluee focuses instead on a special case,  
\textbf{ASV-constrained} t-DCF, where the only information known about the ASV system (a black-box) are the three error rates $P_\text{miss}^\text{asv} \equiv P_\text{miss}^\text{asv}(\tau_{\text{asv}})$,  $P_\text{fa}^\text{asv} \equiv P_\text{fa}^\text{asv}(\tau_{\text{asv}})$ and $P_\text{miss,spoof}^\text{asv} \equiv P_\text{miss,spoof}^\text{asv}(\tau_{\text{asv}})$ provided by another party (\emph{e.g.}, ASV vendor or challenge organizer).

As the main difference between the two t-DCF variants is whether we can adjust the ASV threshold or not, we use the overloaded notation $\text{t-DCF}(\tau_{\text{cm}})$ to  indicate the ASV-constrained t-DCF. With straightforward manipulation, the t-DCF expression of~\eqref{eq:tDCF-unconstrained} can then be rewritten as:
\begin{mdframed}[style=MyFrame]
    \center{\textbf{ASV-constrained t-DCF}}
    \begin{equation}\label{eq:tDCF-ASV-constrained}
        \text{t-DCF}(\tau_\text{cm}) =  C_0+C_1P_\text{miss}^\text{cm}(\tau_\text{cm})
        +C_2P_\text{fa}^\text{cm}(\tau_\text{cm}),
    \end{equation} 
\end{mdframed}    
where $C_0$, $C_1$, and $C_2$ are constants dictated both by the t-DCF parameters and the ASV error rates.  They are given by:
    \begin{equation}\label{eq:tdcf-coefficients}
        \begin{aligned}
            C_0 & = \pi_\text{tar}C_\text{miss}P_\text{miss}^\text{asv}+\pi_\text{non}C_\text{fa}P_\text{fa}^\text{asv}\\
            C_1 & = \pi_\text{tar}C_\text{miss} - \left(\pi_\text{tar}C_\text{miss}P_\text{miss}^\text{asv}+\pi_\text{non}C_\text{fa}P_\text{fa}^\text{asv} \right) \\
            C_2 & = \pi_\text{spoof}C_\text{fa,spoof}
            P_\text{fa,spoof}^\text{asv}.
       \end{aligned}
    \end{equation}
We present an analysis of these coefficients in detail below. First, however, we explain the necessity to \emph{normalize} the raw t-DCF values (whether unconstrained or ASV-constrained case).

\section{Normalized and minimum t-DCF}

Both the individual \eqref{eq:asv-cm-detection-errors} and the tandem \eqref{eq:tandem-error-probabilities} error rates take values in $[0,1]$. This is not the case for the t-DCF in \eqref{eq:tDCF-unconstrained}, however, which is a linear combination of the tandem errors formed by non-negative but otherwise unconstrained multipliers (the products of costs and priors). The `raw' t-DCF values can hence be difficult to interpret, especially across different t-DCF parametrizations. Normalization is performed differently depending on whether one focuses on the unconstrained case (both CM and ASV systems are adjustable) or the ASV-constrained case (only the CM system is adjustable). Let us first focus on the unconstrained case.

\subsection{Normalizing the Unconstrained t-DCF}

Following the practice adopted in the NIST SREs \cite{Greenberg2020_twenty_years,Doddington2000-NIST-overview}, it is preferable to report the \emph{normalized} t-DCF given by:
    \begin{equation}\label{eq:normalized-t-DCF}
        \text{t-DCF}'(\tau_\text{cm},\tau_\text{asv})=\frac{ \text{t-DCF}(\tau_\text{cm},\tau_\text{asv})}{\text{t-DCF}_\text{default}^\text{unconstr}},
    \end{equation}
where $\text{t-DCF}'$ denotes the normalized cost and $\text{t-DCF}_\text{default}^\text{unconstr}>0$ is the t-DCF of a \emph{default} (\emph{reference}) system that yields a fixed decision regardless of input data. The default system either accepts every user, or rejects every user. As an intuitive analogy, the reader may picture a door lock that will either open with any key (inluding that of a burglar), or with no key in the world (including the owner's key). Any useful lock should do better work than either one of these two default options. 

In a similar vein, any useful tandem system should yield a lower cost than that of \emph{both} `accept all' and `reject all' default systems. The former corresponds to action $\alpha_2$ in~\eqref{eq:possible-actions}. When both thresholds are set to $-\infty$, the two miss rates $P_{\text{miss}}^{\text{asv}}(\tau_\text{asv})$ and $P_{\text{miss}}^{\text{cm}}(\tau_\text{cm})$, and hence also $P_\text{a}(\tau_\text{cm},\tau_\text{asv})$ and $P_\text{d}(\tau_\text{cm},\tau_\text{asv})$ in~(\ref{eq:tandem-error-probabilities}) all reduce to zero. $P_\text{b}(\tau_\text{cm},\tau_\text{asv})$ and $P_\text{c}(\tau_\text{cm},\tau_\text{asv})$ reduce to one, giving: 
    \begin{equation}\label{eq:default_alpha2}
        \begin{aligned}
        \text{t-DCF}_{\alpha_2} & = \text{t-DCF}(-\infty,-\infty)\\
        & = C_\text{fa}\cdot \pi_\text{non} + C_\text{fa,spoof}\cdot \pi_\text{spoof},
        \end{aligned}
    \end{equation}
in which there are no target speaker parameters (as there are no misses). In similar fashion and depending on whether 
each trial is rejected by the ASV system (action $\alpha_1$) or by the CM (action $\alpha_3$), the `reject all' default systems are given by:
    \begin{equation}\label{eq:default_alpha1_alpha3}
        \begin{aligned}
            \text{t-DCF}_{\alpha_1} & =\text{t-DCF}(-\infty,\infty)=C_\text{miss}\cdot\pi_\text{tar}\\
            \text{t-DCF}_{\alpha_3} & =\text{t-DCF}(\infty,\tau_\text{asv})=C_\text{miss}\cdot\pi_\text{tar}, \,\,\,\,\,\, \forall\, \tau_\text{asv} \in \mathbb{R}
        \end{aligned}
    \end{equation}
which contains neither nontarget nor spoof terms (both types of trials are correctly rejected). The equality $\text{t-DCF}_{\alpha_1}=\text{t-DCF}_{\alpha_3}$, in turn, reinforces the idea that it does not matter whether it was the CM or the ASV which rejected the target --- it was rejected, and in both cases the user experiences the same inconvenience. 

A useful tandem system should have lower t-DCF than both of the dummy systems \eqref{eq:default_alpha2} and \eqref{eq:default_alpha1_alpha3}. That is, it should yield a cost lower than their \emph{minimum}. The default system is hence chosen according to:
    \begin{equation}\label{eq:tDCF-default-unconstrained}
        \begin{aligned}
        \text{t-DCF}_\text{default}^\text{unconstr} & = \min\,\{\text{t-DCF}_{\alpha_1},\text{t-DCF}_{\alpha_2},\text{t-DCF}_{\alpha_3} \}\\
        &=\min\,\{\text{t-DCF}_{\alpha_1},\text{t-DCF}_{\alpha_2}\}\\
        &=\min\,\{C_\text{fa}\cdot \pi_\text{non} + C_\text{fa,spoof}\cdot \pi_\text{spoof},C_\text{miss}\cdot\pi_\text{tar}\},
        \end{aligned}
    \end{equation}
where the second line follows from \eqref{eq:default_alpha1_alpha3}. Note, however, that the normalized t-DCF obtained by dividing \eqref{eq:tDCF-unconstrained} by \eqref{eq:tDCF-default-unconstrained} is \emph{not} an upper bound. With poorly set detection thresholds (alternatively, using Bayes-optimal thresholds but with badly calibrated scores \cite{LeeuwenB07}), the normalized cost can exceed 1; it can be higher than that of the default system. Such systems are said to be \emph{badly-calibrated}.

An \textbf{optimally calibrated} system provides another useful reference. This \emph{minimum} t-DCF is defined as the minimum cost over all thresholds $(t_{\text{cm}},t_{\text{asv}}) \in \mathbb{R}^2$,
    \begin{equation}
        \text{t-DCF}_\text{min}=\inf_{(t_{\text{cm}},t_{\text{asv}})} \text{t-DCF}(t_{\text{cm}},t_{\text{asv}}),
    \end{equation}
where the infimum (greatest lower bound) is replaced by $\min$ for finite score sets. By definition, $\text{t-DCF}(\tau_{\text{cm}},\tau_{\text{asv}}) \geq \text{t-DCF}_\text{min}$ for any choice of the thresholds (including those of the `default' tandem system). Thus, the normalized minimum cost $\text{t-DCF}'_\text{min}$, is upper bounded by unity:
    \begin{equation}
        \text{t-DCF}'_\text{min}=\frac{\text{t-DCF}_\text{min}}{\text{t-DCF}_\text{default}} \leq\,\, \frac{\text{t-DCF}_\text{min}}{\text{t-DCF}_\text{min}}=1,
    \end{equation}
making it a convenient  number between 0 and 1. 
Like the EER metric, the minimum t-DCF uses an oracle threshold determined with use of ground-truth labels (trial key). 

\begin{table}[!t]\caption{Summary of t-DCF variants and their normalizations. Normalized t-DCF value larger than 1 indicates badly calibrated systems.}\label{tab:tDCF-variant-summary}
\begin{centering}
    \begin{tabular}{c|c|c|c|}
        Type of t-DCF & Raw form & Normalized form & Min. value\\
        \hline\hline
        Unconstrained & \eqref{eq:tDCF-unconstrained} & \eqref{eq:tDCF-unconstrained}$/$\eqref{eq:tDCF-default-unconstrained} & 0\\
        ASV-constrained & \eqref{eq:tDCF-ASV-constrained} &  \eqref{eq:tDCF-ASV-constrained}$/$\eqref{eq:tDCF-default-ASV-constrained} & $C_0$\\
        \hline
    \end{tabular}
\end{centering}
\end{table}

\begin{figure}[!t]
	\centering
  \includegraphics[width=1.00\columnwidth]{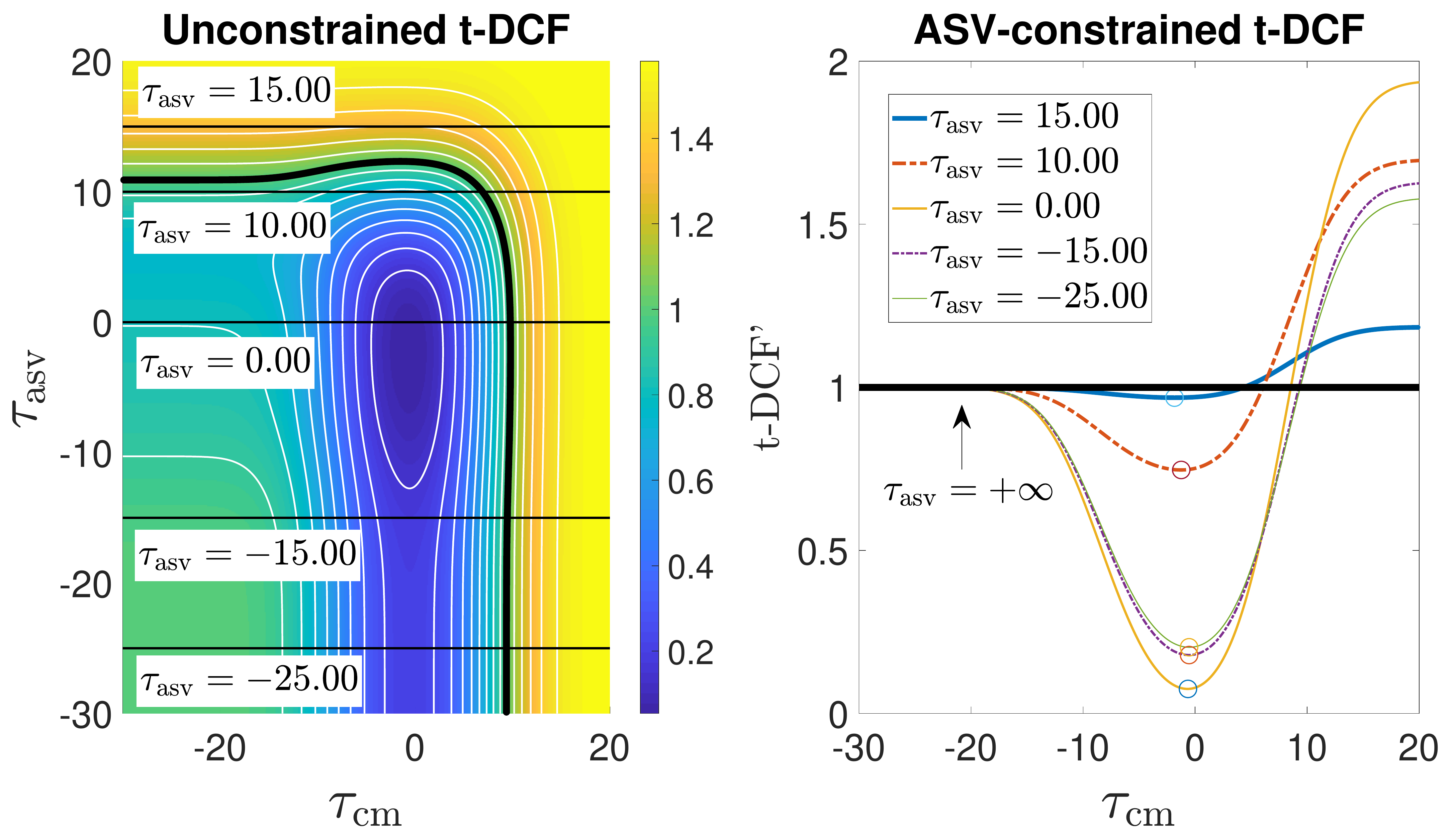}
	\caption{Illustration of unconstrained (left) and ASV-constrained t-DCFs (right) using simulated ASV and CM scores. The former involves both thresholds, while the latter considers `slices' defined by a fixed ASV operating point. Both variants are displayed in their normalized forms (See Table \ref{tab:tDCF-variant-summary}). The isocontour of $\text{t-DCF}'=1$ is highlighted in each case.}
	\label{fig:unconstrained-vs-conditional-tDCF}
\end{figure}

\subsection{Normalizing the ASV-Constrained t-DCF}

A normalised version of the ASV-constrained t-DCF, $\text{t-DCF}(\tau_\text{cm})'=\text{t-DCF}(\tau_\text{cm})/\text{t-DCF}_\text{default}^\text{constr}$, can similarly be defined by selecting an appropriate default cost $\text{t-DCF}_\text{default}^\text{constr}>0$. Since the ASV system is now fixed, the default cost is obtained by adjusting the CM threshold only, with either $\tau_{\text{cm}}=-\infty$ (accept all) or $\tau_{\text{cm}}=+\infty$ (reject all) in \eqref{eq:tDCF-ASV-constrained}, giving:
\begin{equation}
        \begin{aligned}
            \text{t-DCF}_\text{default}^\text{constr}& =\min\{ C_0 + C_1, C_0+C_2\}\\
            & = C_0 + \min\{C_1,C_2\}.
        \end{aligned}            
        \label{eq:tDCF-default-ASV-constrained}
    \end{equation}

\subsection{A Summary of the Two t-DCF Variants}

A summary of the two different t-DCF versions discussed above is given in Table \ref{tab:tDCF-variant-summary}. For the unconstrained case, cost scaling is specified by the t-DCF parameters only. For the ASV-constrained case, it is also dependent on the known, fixed ASV error rates. In both cases, however, normalized costs larger than 1 indicate that the system under consideration cannot do better than `no system'. The `system' differs between the two cases: for the unconstrained case, it is the (CM, ASV) tandem; for the ASV-constrained case, it is the CM only. 

Both t-DCF variants are illustrated in Fig. \ref{fig:unconstrained-vs-conditional-tDCF} for simulated scores (see Appendix) with $P_\text{e}^\text{asv}=0.01$ (ASV EER=1\%), $P_\text{e}^\text{cm}=0.02$ (CM EER=2\%) and $\xi=0.85$ (a parameter defined in Eq. \eqref{eq:appendix-pR_spoof} to model efficacy of spoofing attacks. The arbitrary value 0.85 used here is a proxy of highly effective spoofing attack, see Fig. \ref{fig:spoofing-factor}). The t-DCF parameters are set as described in Section \ref{sec:experimental-setup} (with $\pi_\text{spoof}=0.05$). Even in this idealized simulation, the resulting unconstrained t-DCF has a relatively complicated shape. We observe a valley near the origin $(0,0)$ in the left panel of Fig. \ref{fig:unconstrained-vs-conditional-tDCF}. Moving away from this `sweet spot' yields increased costs. Whenever either threshold is too high/low, we approach the `accept all' or `reject all' cases which are generally suboptimal. 

Focusing on the ASV-conditional case and comparing the t-DCF values across five arbitrary ASV operating points indicated in Fig.~\ref{fig:unconstrained-vs-conditional-tDCF}, the lowest min t-DCF is obtained for $\tau_\text{asv}=0.00$ (which coincides with the EER operating point in our simulation). We also observe that (a) the t-DCF function flattens with increasing $\tau_\text{asv}$, and (b) it reaches the value of 1 at one of the infinities (here, at $\tau_\text{cm}=-\infty$). These two properties hold for any ASV and CM system, and will be elaborated on below.

\section{Analysis of the ASV-Constrained t-DCF}

We now provide some intepretation of the coefficients in \eqref{eq:tdcf-coefficients}.
First, the offset $C_0 \geq 0$ is dubbed the \emph{ASV floor} as it lower bounds \eqref{eq:tDCF-ASV-constrained} and vanishes only for a perfect ASV system ($P_\text{miss}^\text{asv}=P_\text{fa}^\text{asv}=0$). Note that, even though $C_0$ resembles the NIST DCF \eqref{eq:NIST-DCF}, they are not the same; unlike for the NIST DCF, $\pi_\text{tar}+\pi_\text{non} \neq 1$ since some probability mass is assigned to $\pi_\text{spoof}$.   
$C_1$ and $C_2$ reflect the relative importance of the CM miss and false alarm rates in \eqref{eq:tDCF-ASV-constrained}. 

The coefficients in \eqref{eq:tdcf-coefficients} may seem complicated. Some insight into their influence on the t-DCF can be gained by setting the ASV system to the EER operating point so that $ P_\text{miss}^\text{asv}=P_\text{fa}^\text{asv}\equiv P_\text{e}^\text{asv}$, where $P_\text{e}^\text{asv}$ indicates the EER of the ASV system. The choice of EER operating point may look arbitrary as it contradicts the t-DCF parameter specifications. Nonetheless, the EER serves as \emph{tight upper bound} on the binary classifier Bayes error-rate~\cite[p.72]{Brummer2010-PhD}. By fixing the ASV system to the EER operating point, we mimic a miscalibrated ASV system which operates with the \emph{worst possible} target-nontarget discrimination performance. 

The coefficients in (\ref{eq:tdcf-coefficients}) can now be rewritten as functions of either the EER or the SFAR of the ASV system:
    \begin{equation}\label{eq:tdcf-coefficients-simplified}
        \begin{aligned}
            C_0(P_\text{e}^\text{asv}) & = \alpha P_\text{e}^\text{asv}\\
            C_1(P_\text{e}^\text{asv}) & = \beta - \alpha P_\text{e}^\text{asv}\\
            C_2(P_\text{fa,spoof}^\text{asv}) & = \gamma P_\text{fa,spoof}^\text{asv},
        \end{aligned}
    \end{equation}
where $\alpha=\pi_\text{tar}C_\text{miss}+\pi_\text{non}C_\text{fa}$, $\beta=\pi_\text{tar}C_\text{miss}$, and $\gamma=\pi_\text{spoof}C_\text{fa,spoof}$ are constants. By substituting \eqref{eq:tdcf-coefficients-simplified} to \eqref{eq:tDCF-ASV-constrained} we obtain: 
    \begin{equation}
        \begin{aligned}\label{eq:tDCF-ASV-constrained-2}
        \text{t-DCF}(\tau_\text{cm}) = \alpha P_\text{e}^\text{asv}+\left(\beta-\alpha P_\text{e}^\text{asv} \right)P_\text{miss}^\text{cm}(\tau_\text{cm})\\
        +\gamma P_\text{fa,spoof}^\text{asv}P_\text{fa}^\text{cm}(\tau_\text{cm}).  
        \end{aligned}
    \end{equation}   
The influence upon the t-DCF, or weight of CM misses ($C_1$) is a function of the ASV system accuracy (encoded in $P_\text{e}^\text{asv}$).  The weight of CM false alarms ($C_2$) is a function of the ASV system sensitivity to spoofing attacks (encoded in  $P_\text{fa,spoof}^\text{asv}$). Once the evaluation conditions and the performance of the unprotected ASV system in the same conditions is known, then the CM may be optimised using the t-DCF metric (\ref{eq:tDCF-ASV-constrained-2}) tailored to the specific ASV system and evaluation conditions. Here we are not concerned \emph{how} such optimization (involving generally non-differentiable functions due to hard error counting) should be performed --- we are merely stating the objective.

\begin{figure*}
\subfigure[Simulated ASV scores.]{\includegraphics[width=.5\textwidth]{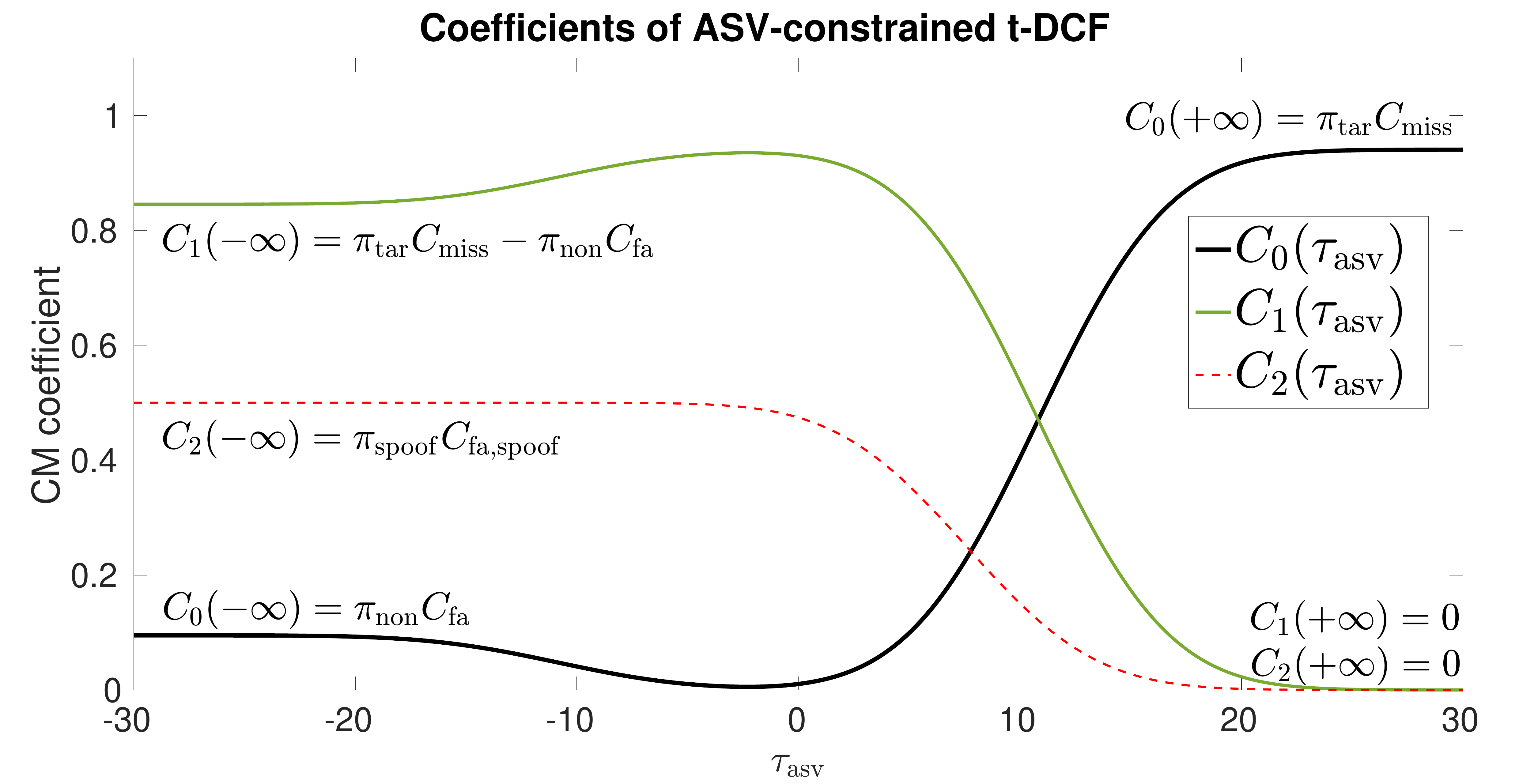}}
\subfigure[x-vector ASV scores.]{\includegraphics[width=.5\textwidth]{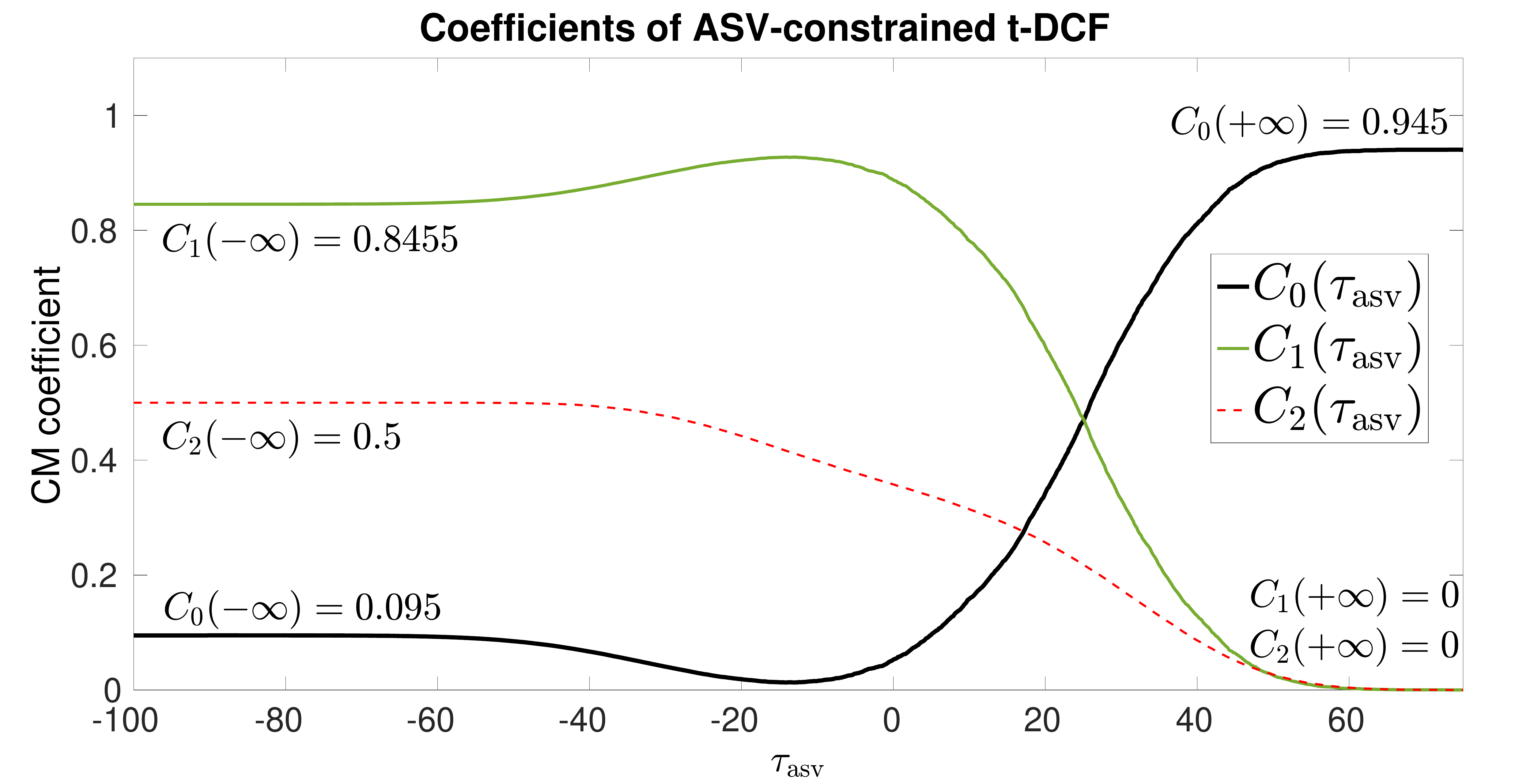}}
	\caption{Illustration of ASV-constrained t-DCF coefficients \eqref{eq:tdcf-coefficients} for (a) simulated ASV scores ($P_\text{e}^\text{asv}=0.01$, $\xi=0.85$) described in the Appendix, and (b) real x-vector ASV scores computed on the ASVspoof 2019 LA database, with t-DCF parameters described in Section \ref{sec:experimental-setup} (with $\pi_\text{spoof}=0.05$). The indicated special values indicate the coefficient values at $\tau_\text{asv}=\pm \infty$. See text for details.}
	\label{fig:asv-cond-coeffs}
\end{figure*}

Another way to analyze $C_0$, $C_1$ and $C_2$ is directly in terms of the ASV threshold. This is visualized in Fig. \ref{fig:asv-cond-coeffs} both for simulated, Gaussian-distributed ASV scores (see Appendix) and real x-vector based ASV scores (see Section \ref{sec:experimental-setup}). The data used for drawing the graphs in the two panels are unrelated; the resemblance of simulated and real functions is indicative of their general shape. Also illustrated are the limits of each coefficient as $\tau_\text{asv} \rightarrow \pm \infty$. As seen, the coefficients are nonlinear functions of the ASV operating point. The  `$\mathcal{X}$' shape formed by $C_0$ and $C_1$ is explained by the dependence $C_0(\tau_\text{asv})+C_1(\tau_\text{asv})=\pi_\text{tar}C_\text{miss}$ seen from Eq. \eqref{eq:tdcf-coefficients}. We now have the following interpretations:
    \begin{itemize}
        \item $C_0$ approaches the `accept all' and `reject all' dummy systems at $\tau_\text{asv}=-\infty$ and $\tau_\text{asv}=+\infty$, respectively. What remains in the ASV floor is either the nontarget ($\pi_\text{non}C_\text{fa}$) or the target ($\pi_\text{tar}C_\text{miss}$) term.
        \item $C_1$ at $\tau_\text{asv}=-\infty$ leads to a curious subtractive expression (which can also take negative values), $\pi_\text{tar}C_\text{miss} - \pi_\text{non}C_\text{fa}$. How so? The dummy ASV system is set to `accept all'. The CM nonetheless offers potential to reject some trials. Subtraction of the nonnegative nontarget term \emph{lowers} $C_1$, which in turn encourages a \emph{higher} CM threshold. For $\tau_\text{asv}\rightarrow+\infty$, $C_1$ vanishes. Since the ASV system will always reject target trials, CM behaviour is irrelevant.
        \item $C_2$ is a nonincreasing function of $\tau_\text{asv}$. For the `accept all' ASV at $\tau_\text{asv}\rightarrow -\infty$, the CM has the tightest security (highest $C_2$). Similar to $C_1$, $C_2$ vanishes at $\tau_\text{asv}\rightarrow+\infty$: such ASV rejects also the spoofing attacks, so the spoof false alarms of CM does not matter.
    \end{itemize}

Concerning the `flattening' of the ASV-conditional t-DCF observed in Fig.~\ref{fig:unconstrained-vs-conditional-tDCF}, note that for the `reject all' case ($\tau_\text{asv}\rightarrow +\infty)$ we have
    \begin{equation}
        \begin{aligned}
        \text{t-DCF}'(\tau_\text{cm})& =\frac{C_0 + C_1P_\text{miss}^\text{cm}(\tau_\text{cm})+C_2P_\text{fa}^\text{cm}(\tau_\text{cm})}{C_0 + \min\{C_1,C_2\}} \nonumber\\
        & = \frac{C_0 + 0\cdot P_\text{miss}^\text{cm}(\tau_\text{cm})+0\cdot P_\text{fa}^\text{cm}(\tau_\text{cm})}{C_0 + \min\{0,0\}} = \frac{C_0}{C_0} = 1, 
        \end{aligned}
    \end{equation}
regardless of the CM system or its operating point. This simply says there is no way to improve (or for that matter, to degrade) such an ASV system using \emph{any} CM. For the `accept all' ASV system ($\tau_\text{asv}\rightarrow -\infty)$, the situation is similar. Performance cannot be improved using any CM ($C_1=C_2=0$) if (and only if)
    \begin{equation}
        (\pi_\text{tar}C_\text{miss}=\pi_\text{non}C_\text{fa})\,\,\ \texttt{AND}\,\,\, (\pi_\text{spoof}=0\,\,\, \texttt{OR}\,\,\, C_\text{fa,spoof}=0).\nonumber
    \end{equation}
The first condition above states that overall costs from missed targets and falsely accepted nontargets are the same (there is no preference for either). The second condition states that either spoofing attacks are not anticipated  ($\pi_\text{spoof}=0$), or that one does not care about them ($C_\text{fa,spoof}=0$). This is intuitively reasonable. The CM cannot do anything useful to reject spoofs (they do not incur losses), and the potential benefit of CMs helping to reject nontargets will be `evened out' by equally costly target rejections (that the dummy ASV system would have otherwise accepted). Whenever the t-DCF parameters and the ASV operating point are chosen so that $C_1 \neq 0$ and $C_2 > 0$, there is potential for the CM to improve performance. 

Finally, why does the normalized ASV-conditional t-DCF reach the value 1 at one of the infinities (as the right panel of Fig. \ref{fig:unconstrained-vs-conditional-tDCF} suggests)? Without loss of generality, suppose that $\text{t-DCF}'(+\infty) \neq 1$. Since $P_\text{miss}^\text{cm}(+\infty)=1$ and $P_\text{fa}^\text{cm}(+\infty)=0$, we have 
    \begin{equation}
        \begin{aligned}
        \text{t-DCF}'(+\infty)& =\frac{C_0 + C_1\cdot 1+C_2\cdot 0 }{C_0 + \min\{C_1,C_2\}} = \frac{C_0 + C_1}{C_0 + \min\{C_1,C_2\}} \nonumber, 
        \end{aligned}
    \end{equation}
and since we assumed this expression is $\neq 1$, it follows that $\min\{C_1,C_2\}=C_2$. Therefore, at $\tau_\text{cm}=-\infty$, we have $\text{t-DCF}'(-\infty)=(C_0 + C_2)/(C_0 + \min\{C_1,C_2\})=1$. Similar argumentation can be made by assuming $\text{t-DCF}'(-\infty) \neq 1$, which implies $\text{t-DCF}'(+\infty)=1$. \emph{At either (CM) infinity, the normalized t-DCF equals 1}, as the system collapses to the default system.

\section{Experimental Set-Up}\label{sec:experimental-setup}

The experimental work aims to assess the tandem operation of ASV and CM systems submitted to the three editions of the \emph{Automatic Speaker Verification Spoofing and Countermeasures} (ASVspoof) challenge. This section defines the cost model parameters, gives an overview of the ASVspoof corpora and the fixed ASV system.

We focus on \textbf{authentication} scenarios, to which the problem of spoofing is most relevant. As in  \cite{Kinnunen2018-tDCF}, we assume a hypothetical banking application where $C_\text{miss}=1$, $C_\text{fa}=C_\text{fa,spoof}=10$ and $\pi_\text{non} \ll \pi_\text{tar}$, and $\pi_\text{spoof} \ll \pi_\text{tar}$. The parameter of interest is $\pi_\text{spoof}$, which we fix to a small arbitrary value and then obtain $\pi_\text{tar}=(1 - \pi_\text{spoof})\times 0.99$ and $\pi_\text{non}=(1 - \pi_\text{spoof})\times 0.01$. 

The three speech corpora originate from the past ASVspoof challenges. The 2015 edition \cite{Wu2015-asvspoof} focused on the detection of synthetic speech and voice conversion, the 2017 edition \cite{Kinnunen2017assessing} focused on the detection of replay attacks and the latest 2019 edition \cite{Todisco2019} focused on the three types of attacks categorized into \emph{logical access} (LA) and \emph{physical access} (PA) scenarios. The data and protocol related details of these corpora are reported elsewhere~\cite{Wu2015-asvspoof,Kinnunen2017assessing,Todisco2019}; the focus here is on aspects relevant to the current evaluation. A summary of trial statistics for the evaluation partitions of the corpora used in this work is presented in Table \ref{tab:asvspoof-trial-statistics}.

\begin{table}[t]
	\renewcommand{\arraystretch}{1.2}
	\caption{Number of trials in the evaluation protocols for ASV experiments.}
	\centering
	\vspace{0.1cm}
	\begin{footnotesize}
		\begin{tabular}{|c|c|}
			\hline
			Dataset Name                   & Target / Non-target / Spoof \\
			\hline
			ASVspoof 2015        & 4053 / 77007 / 80000\\
			ASVspoof 2017        & 1106 / 18624 / 10878\\
			ASVspoof 2019 (LA)   & 5370 / 33327 / 63882\\
			ASVspoof 2019 (PA)   & 12960 / 123930 / 116640\\
			\hline
		\end{tabular}
	\end{footnotesize}
	\label{tab:asvspoof-trial-statistics}
\end{table}

The ASV system uses \emph{time-delay neural network} (TDNN) based \emph{x-vector} speaker embeddings \cite{Snyder2018XVectorsRD} together with a \emph{probabilistic linear discriminant analysis} (PLDA) \cite{prince2007probabilistic} backend. The x-vector extractor is a pre-trained\footnote{\url{http://kaldi-asr.org/models/m7}} neural network model developed with the Kaldi~\cite{povey2011kaldi} toolkit. It is trained with MFCC features extracted from audio data from 7,325 speakers of the VoxCeleb1 and VoxCeleb2 corpora~\cite{nagrani2017voxceleb}.\footnote{For further details, check the VoxCeleb Kaldi recipe at \url{https://github.com/kaldi-asr/kaldi/tree/master/egs/voxceleb/v2}} Further details related to network parameters and data preparation are available in~\cite{Snyder2018XVectorsRD}. The original Kaldi recipe was modified to include PLDA adaptation using in-domain data. Full details of the ASV system can be found in \cite{wang2019asvspoof}. PLDA adaptation is applied separately for the ASVspoof 2015, 2017, and 2019 (LA and PA) datasets using in-domain data.

\section{Results}

First, the unconstrained and ASV-constrained variants are compared empirically. Then, the ASV-constrained t-DCF is used to assess the performance of submitted CM systems together with a common ASV system. The ASV threshold $\tau_\text{asv}$ is fixed to the EER operating point, while the CM threshold is set to the minimum point of t-DCF. The final experiment addresses the choice of thresholds. All the presented t-DCF values are in their normalised form (see Table \ref{tab:tDCF-variant-summary}).

\subsection{Unconstrained vs. ASV-Constrained t-DCF}

\begin{figure}[!t]
    \centering
    \includegraphics[width=0.99\columnwidth]{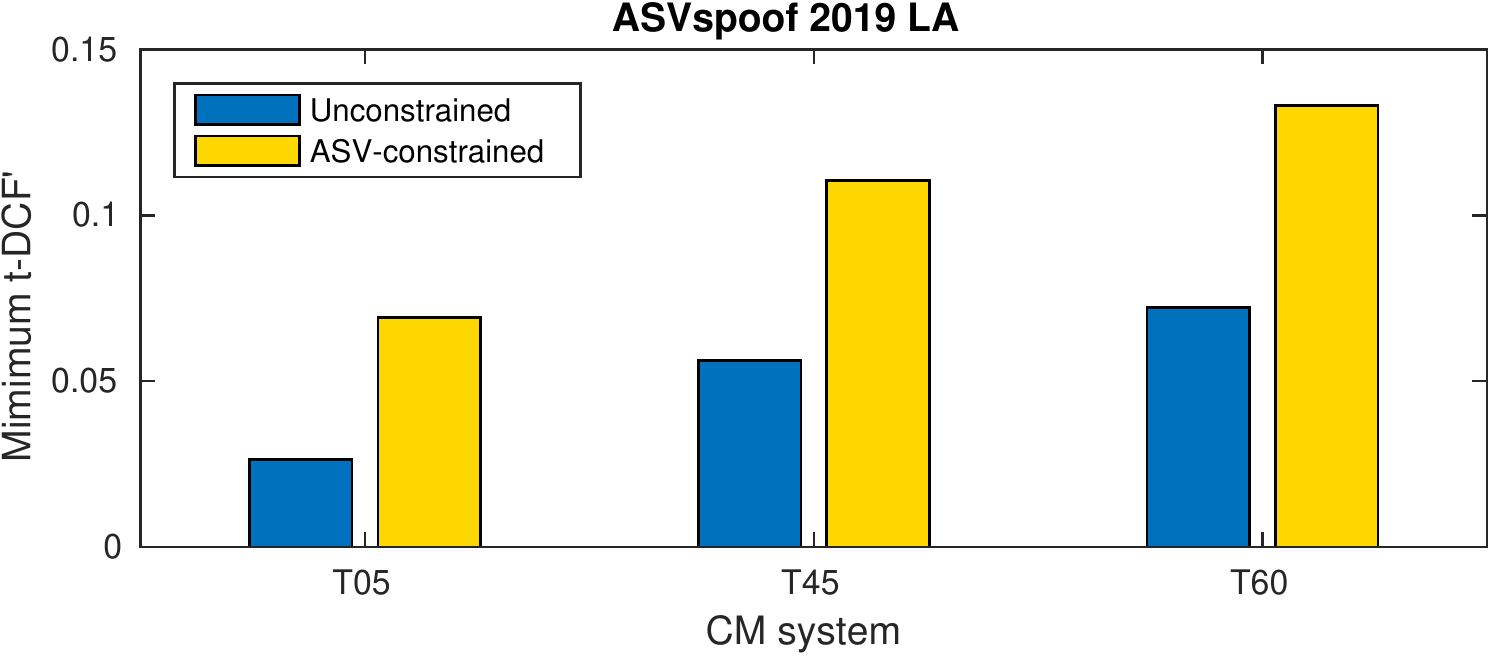}
    \caption{Comparison of the unconstrained and ASV-constrained minimum t-DCF for the top three systems of the ASVspoof 2019 LA challenge ($\pi_\text{spoof}=0.05$). For t-DCF both ASV and CM thresholds are chosen to minimize the cost, while for ASV-constrained t-DCF, the ASV is set to its EER point.}
    \label{fig:tDCF_comparison}
\end{figure}

Fig.~\ref{fig:tDCF_comparison} illustrates a comparison of unconstrained and ASV-constrained t-DCF formulations for the three top-performing systems of the ASVspoof 2019 LA scenario. The ASV-constrained values are systematically higher, as expected: while the unconstrained t-DCF does not assume a pre-defined ASV threshold, the ASV-constrained t-DCF does. The minimum t-DCF for the ASV-constrained is lower bounded by the cost of the ASV system at the EER operating point (which is suboptimal), the unconstrained t-DCF allows both ASV and CM thresholds to be varied jointly, yielding lower t-DCF values. Another difference is in the default cost used to obtain the normalised t-DCF: while the default t-DCF for the unconstrained formulation does not depend on ASV error rates --- see Eqs.~\eqref{eq:default_alpha2}, \eqref{eq:default_alpha1_alpha3} and \eqref{eq:tDCF-default-unconstrained} --- the one for the ASV-constrained t-DCF does; see Eqs.~\eqref{eq:tdcf-coefficients} and \eqref{eq:tDCF-default-ASV-constrained}. 

\subsection{ASV-Constrained t-DCF of ASVspoof Submissions}

\begin{figure*}
    \centering
    \begin{subfigure}[$\pi_\text{spoof}=0.01$.]{
        \includegraphics[width=0.24\textwidth]{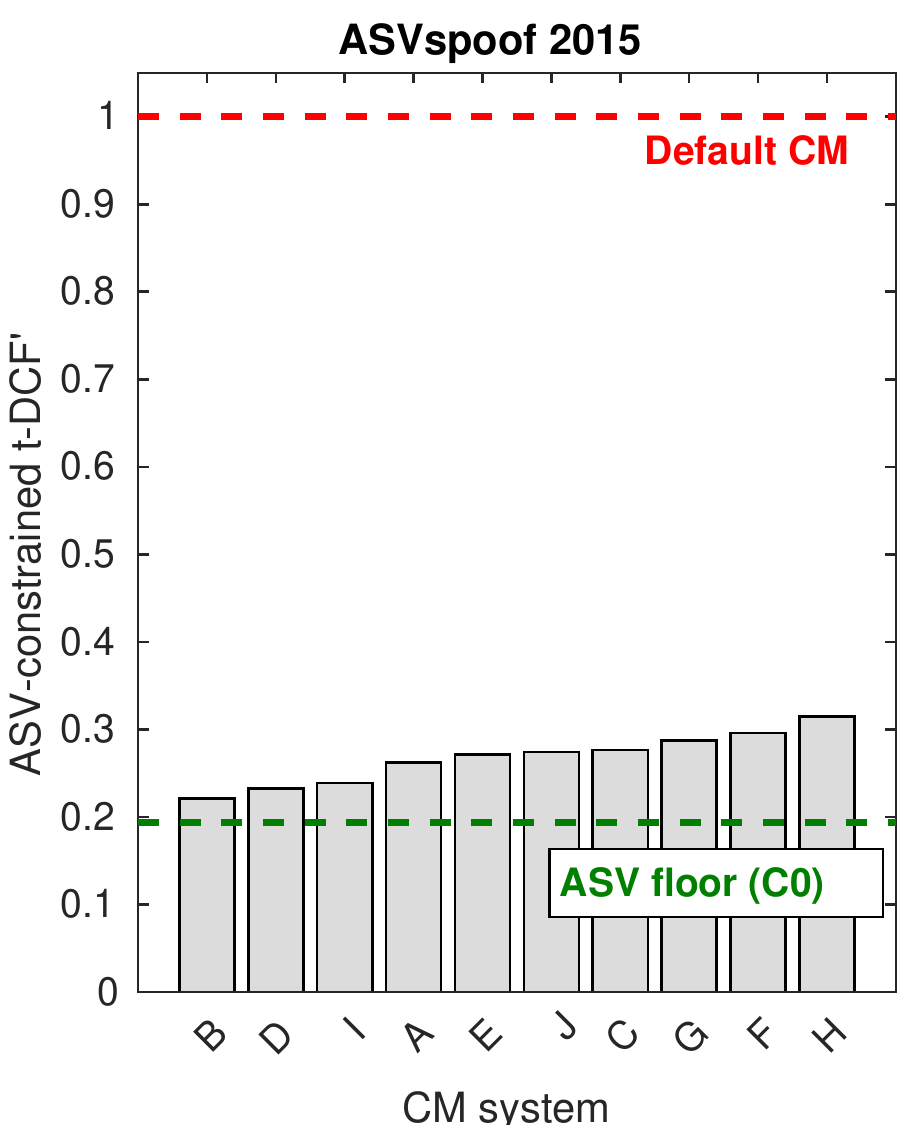}
        \includegraphics[width=0.24\textwidth]{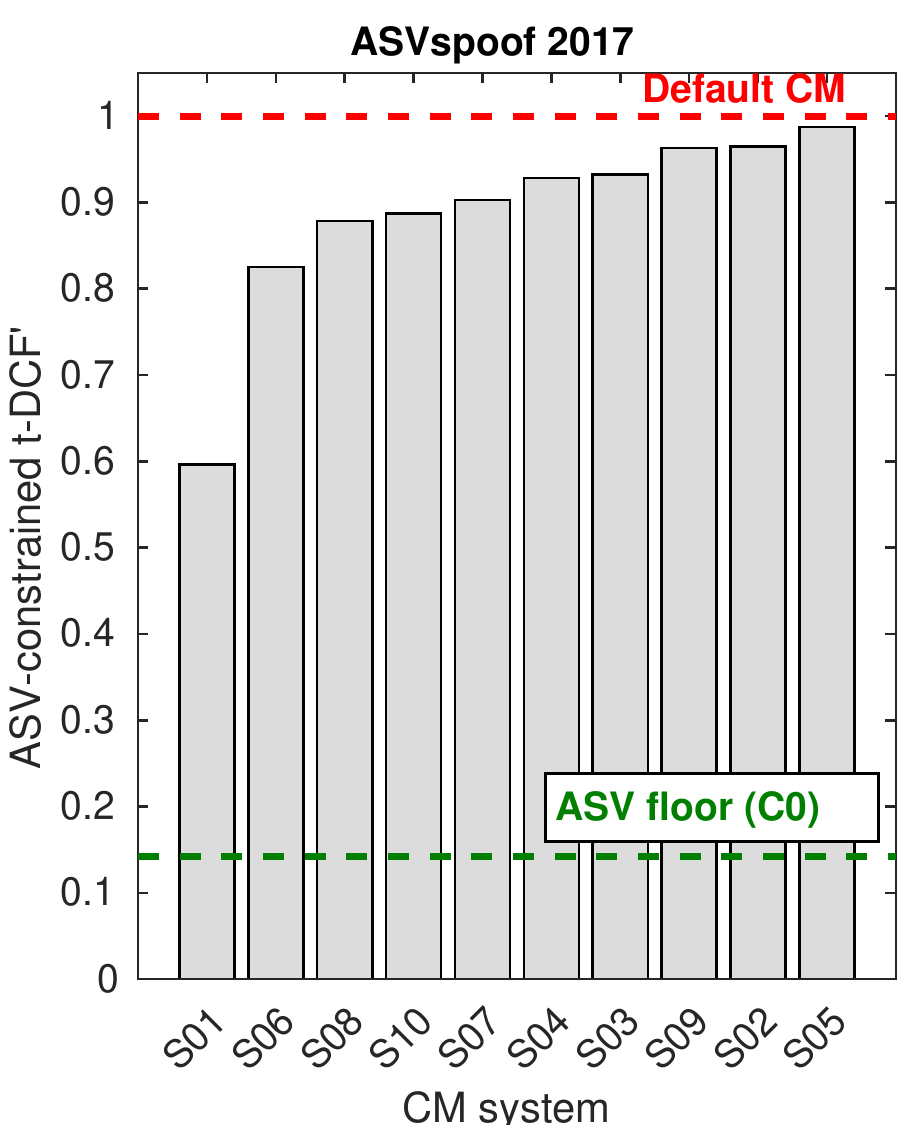}
        \includegraphics[width=0.24\textwidth]{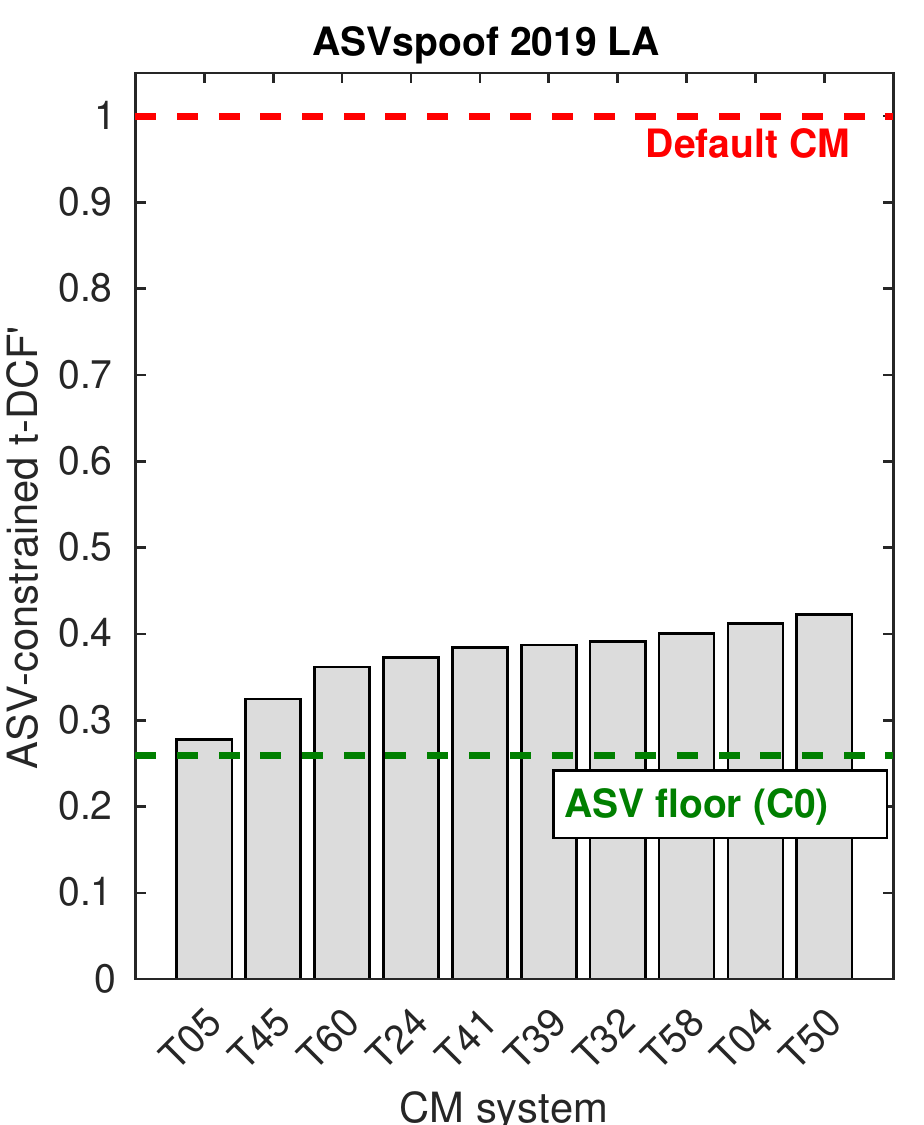}
        \includegraphics[width=0.24\textwidth]{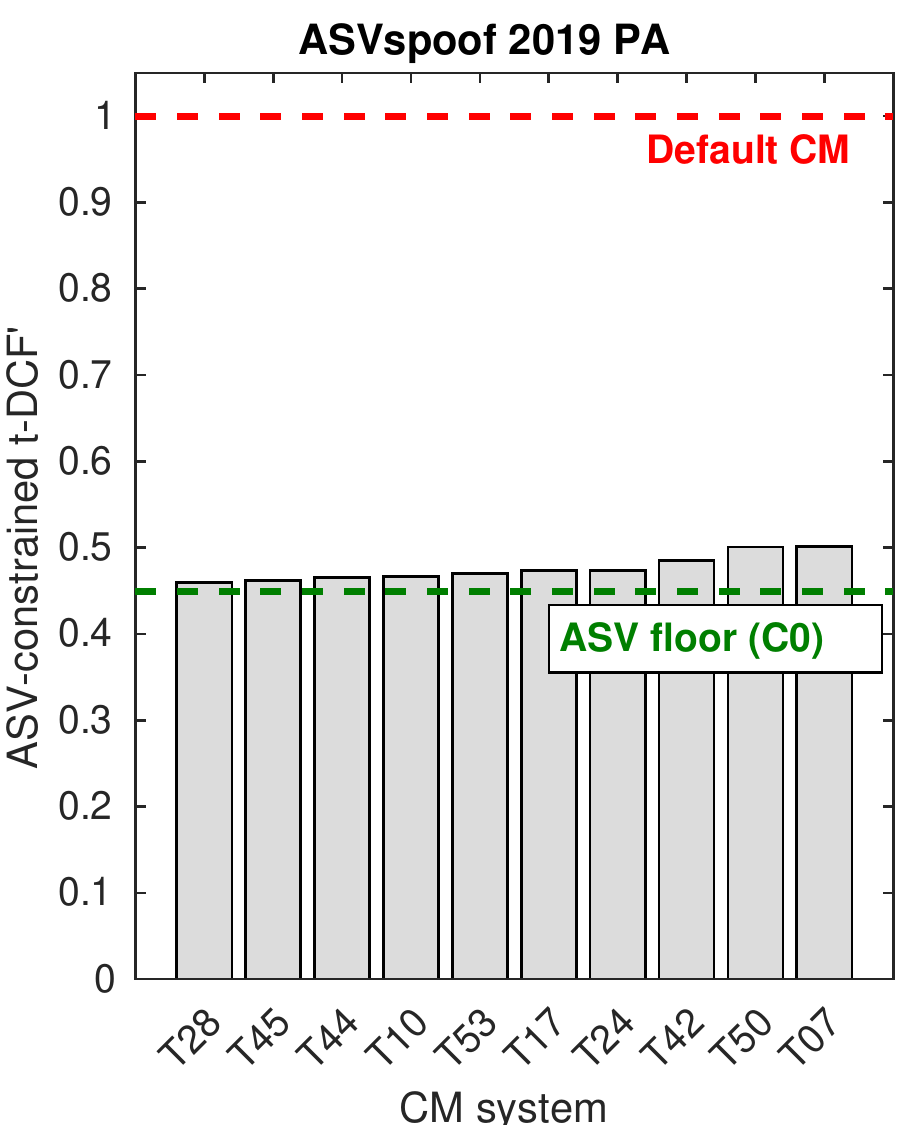}}
    \end{subfigure}
    \begin{subfigure}[$\pi_\text{spoof}=0.05$.]{
        \includegraphics[width=0.24\textwidth]{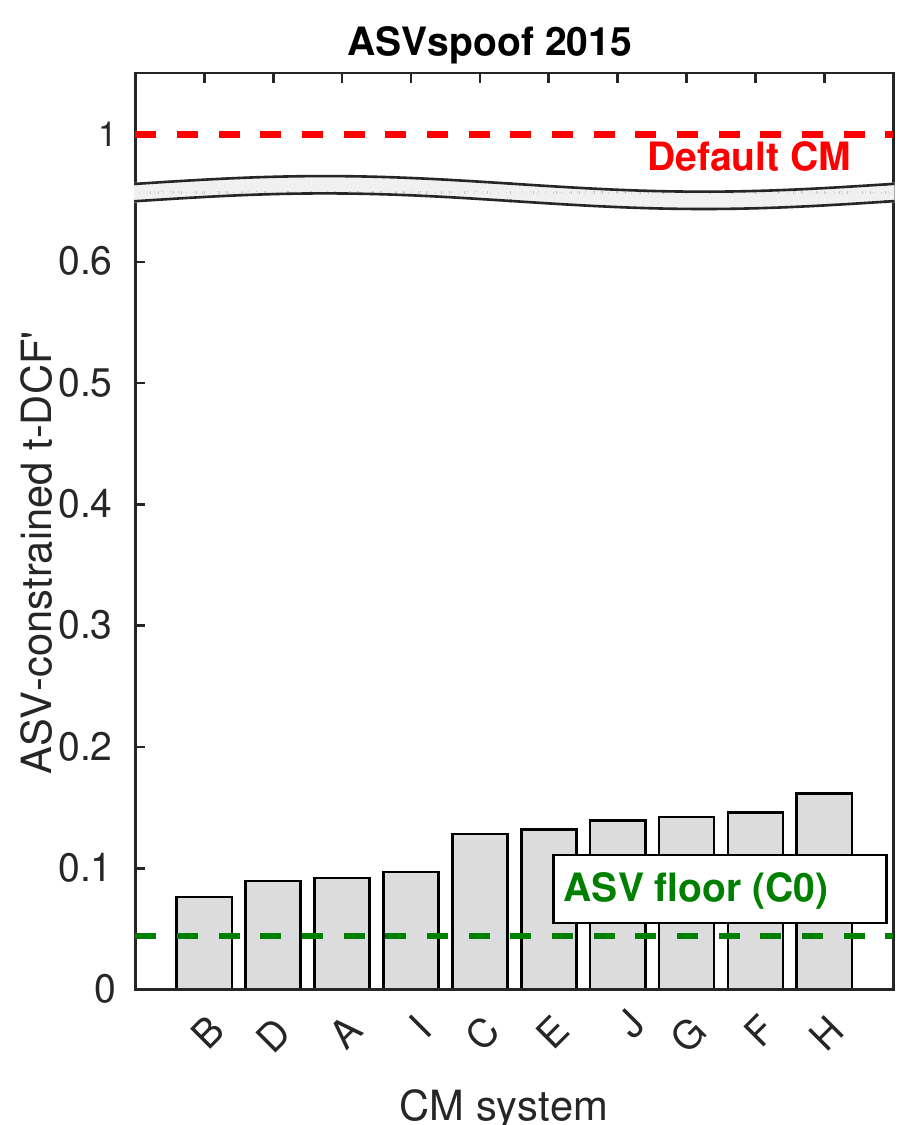}
        \includegraphics[width=0.24\textwidth]{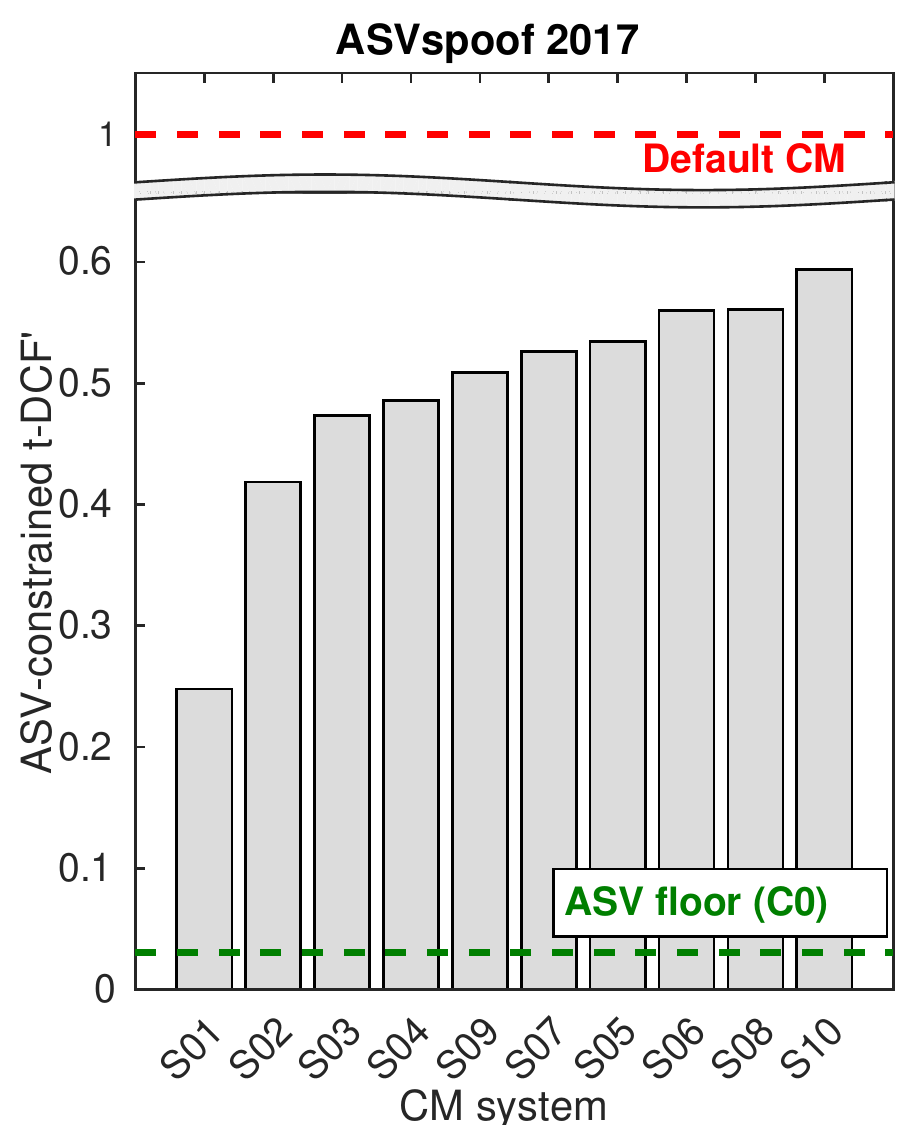}
        \includegraphics[width=0.24\textwidth]{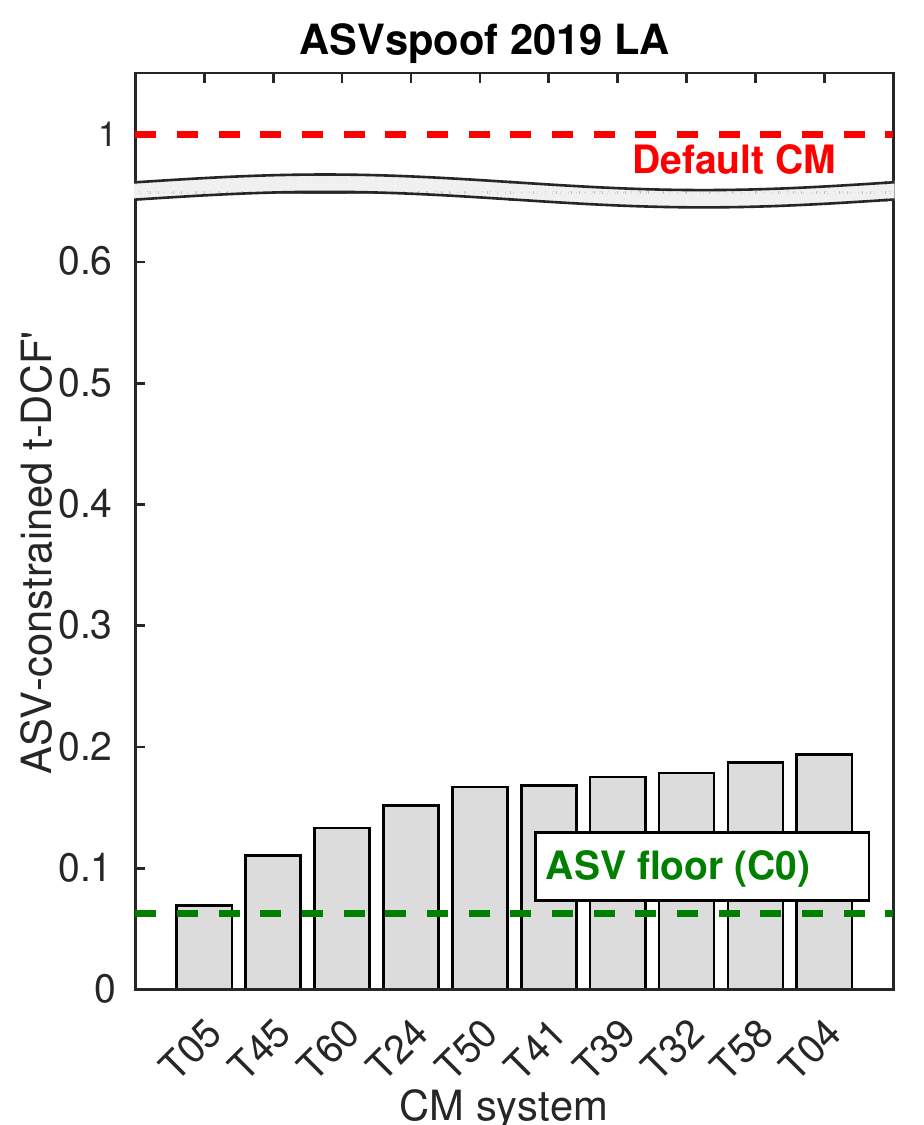}
        \includegraphics[width=0.24\textwidth]{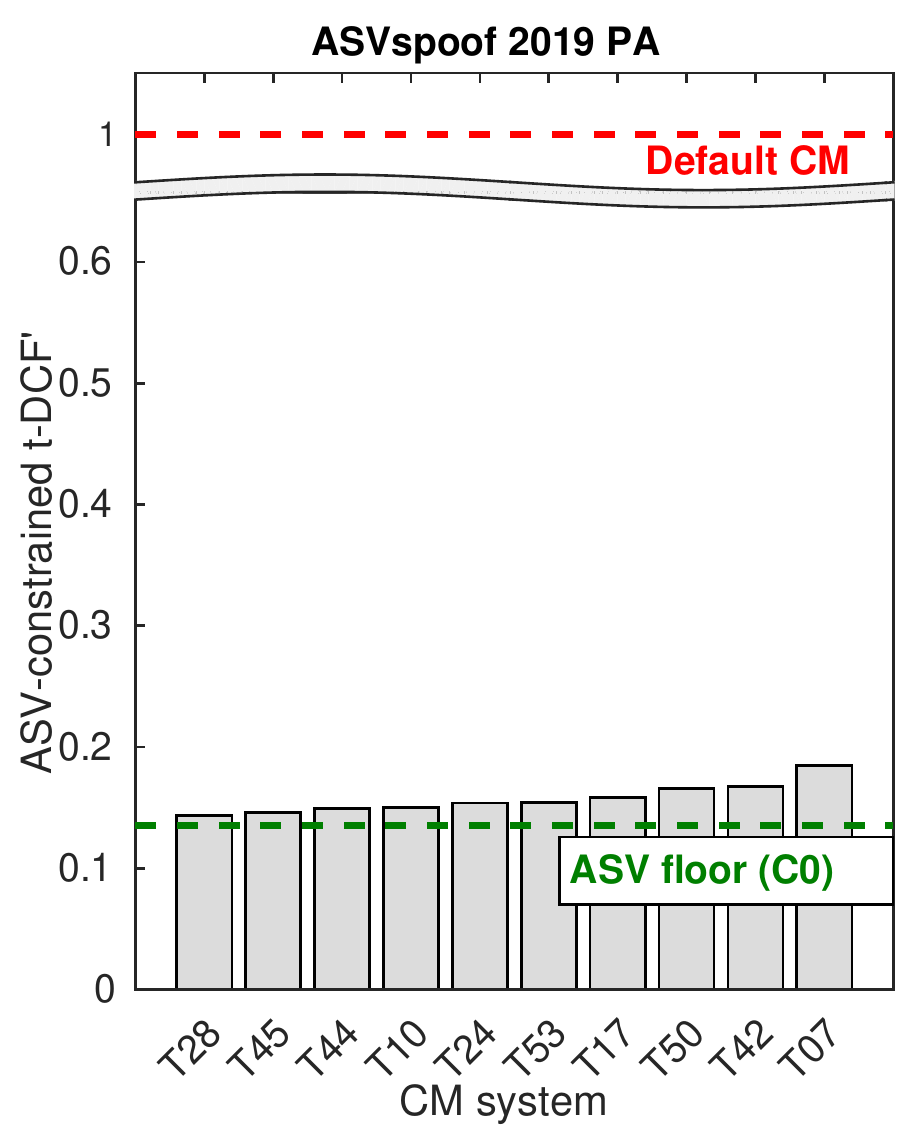}}
    \end{subfigure}
    \caption{ASV-constrained minimum normalised t-DCF of top-10 systems of ASVspoof 2015, 2017, 2019 LA and 2019 PA challenges for (a) $\pi_{\text{spoof}}=0.01$ and (b) $\pi_{\text{spoof}}=0.05$. ``ASV floor" indicates the cost of the tandem system if the CM system was perfect (equal to $C_0$). ``Default CM'' indicates the cost of the tandem system with a dummy CM that either accepts or rejects all the trials (whichever yields a lower cost).} \label{fig:tDCF_allDBs}
\end{figure*}

Fig.~\ref{fig:tDCF_allDBs} illustrates the ASV-constrained, minimum normalised t-DCF of the ten top-performing submission of each ASVspoof challenge (2015, 2017, 2019 LA and 2019 PA). Submissions are sorted by increasing t-DCF. The upper and lower figures were computed using $\pi_\text{spoof}=0.01$ and $\pi_\text{spoof}=0.05$, respectively. For reference purposes, the green line shows the t-DCF for a \emph{perfect} CM. It corresponds to the ASV floor $C_0$ as defined in Eq.~\eqref{eq:tdcf-coefficients}. Another reference, shown by the red dashed line, corresponds to the default CM that either accepts or rejects all trials, whichever produces a lower cost.

Our first two immediate observations are that CMs are beneficial (all values are below `no CM') and that the improvements are often substantial. Second, none of the CMs reaches the ASV floor, suggesting potential for future improvements in the CM technology. We see overall \emph{higher} t-DCF values obtained with \emph{lower} spoofing prior $\pi_\text{spoof}=0.01$. This may seem counterintuitive at first but the operation of an imperfect CM will also produce target speaker misses, which increases the overall cost. In the extreme case when no spoofing attacks are expected, one should not use \emph{any} CM. In contrast, when spoofing attacks are likely to occur, CMs are helpful in decreasing the cost by rejecting spoofing attempts (relative to not having any CM).

Note that the CM rank may change when  $\pi_\text{spoof}$ varies. Differences in the ranks are notable for the ASVspoof 2017 database. This is explained by noting that the target metric for the ASVspoof 2017 edition was the CM EER, which is generally not the optimal operating point for the t-DCF parameters used here. A given system (e.g.~S02) can perform better than others at the EER point, but worse in other areas of the DET curve.  

\begin{figure}
    \centering
    \includegraphics[width=\columnwidth]{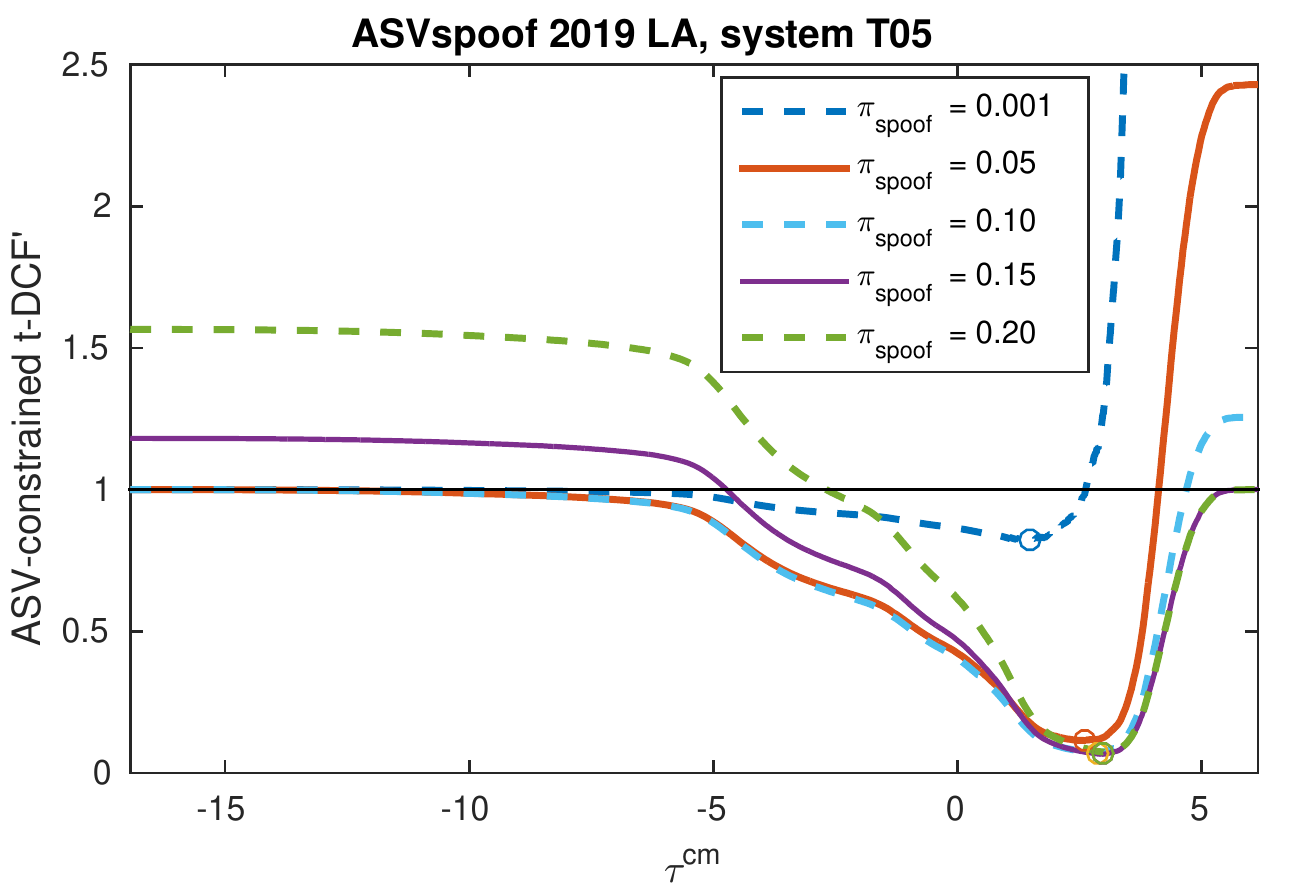}
    \caption{Normalised ASV-constrained t-DCF w.r.t. the CM threshold for system T45 on ASVspoof 2019 LA.}
    \label{fig:LA_T45_tDCF_vs_spoofPrior}
\end{figure}

Fig.~\ref{fig:LA_T45_tDCF_vs_spoofPrior} can give some insights into the observations made from Fig.~\ref{fig:tDCF_allDBs}. Here, the ASV-constrained normalised t-DCF curve is shown for submission T45 of the ASVspoof 2019 LA challenge, when varying the CM threshold $\tau^\text{cm}$, for different values of $\pi_\text{spoof}$: 0.05, 0.10, 0.15 and 0.20. We see that the (minimum) t-DCF decreases with an increasing value of $\pi_\text{spoof}$. The optima are reached for different values of $\tau_\text{cm}$ depending on $\pi_\text{spoof}$, as expected. For the lowest spoof prior $\pi_\text{spoof}=0.001$, minimum t-DCF is only slightly below 1, indicating that the CM cannot improve the performance much; even though T45 performs well, the spoofing attack is simply too rare for it to make a substantial difference.

\begin{figure}[!t]
    \centering
    \includegraphics[width=0.99\columnwidth]{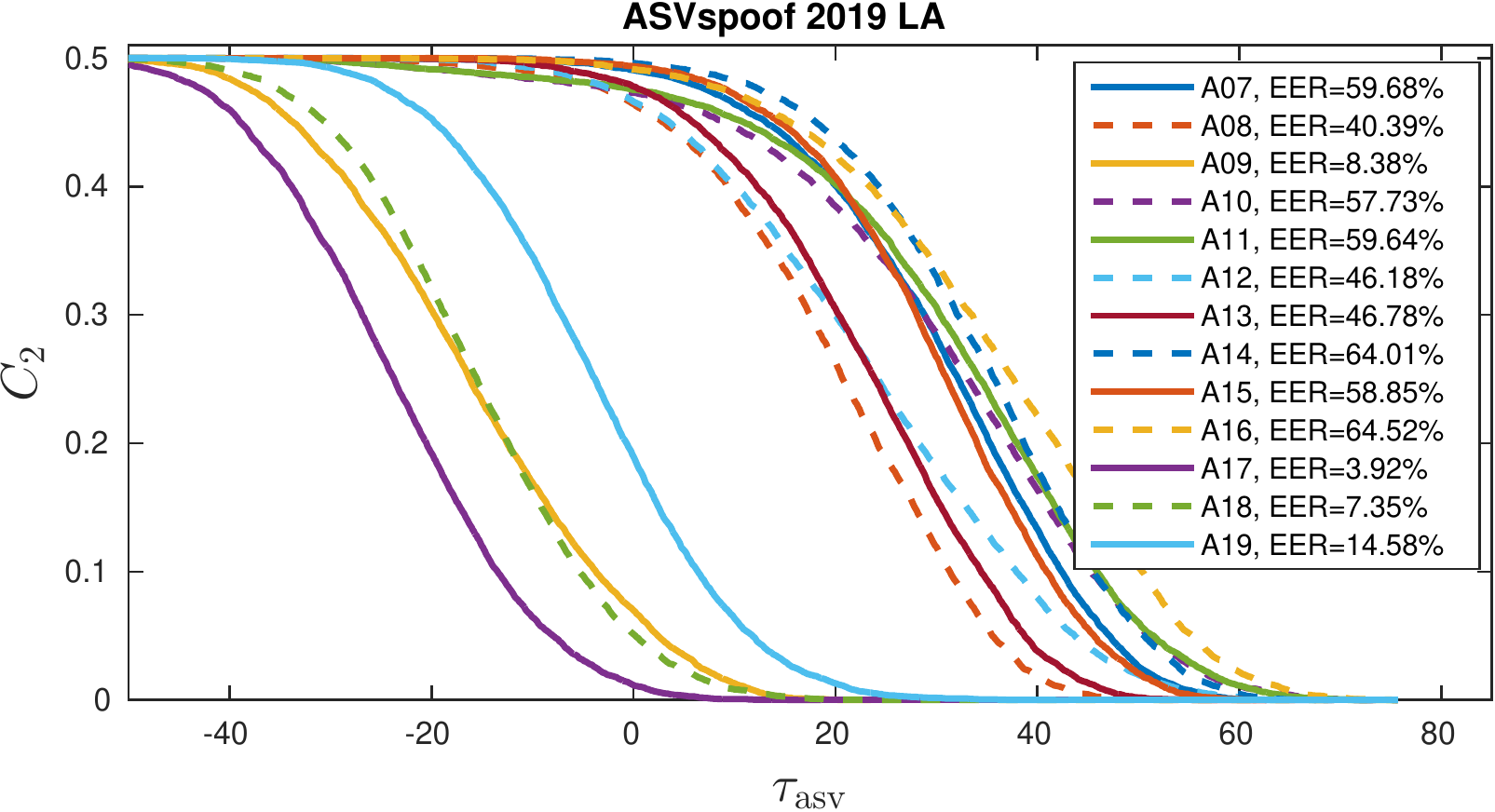}
    \includegraphics[width=0.99\columnwidth]{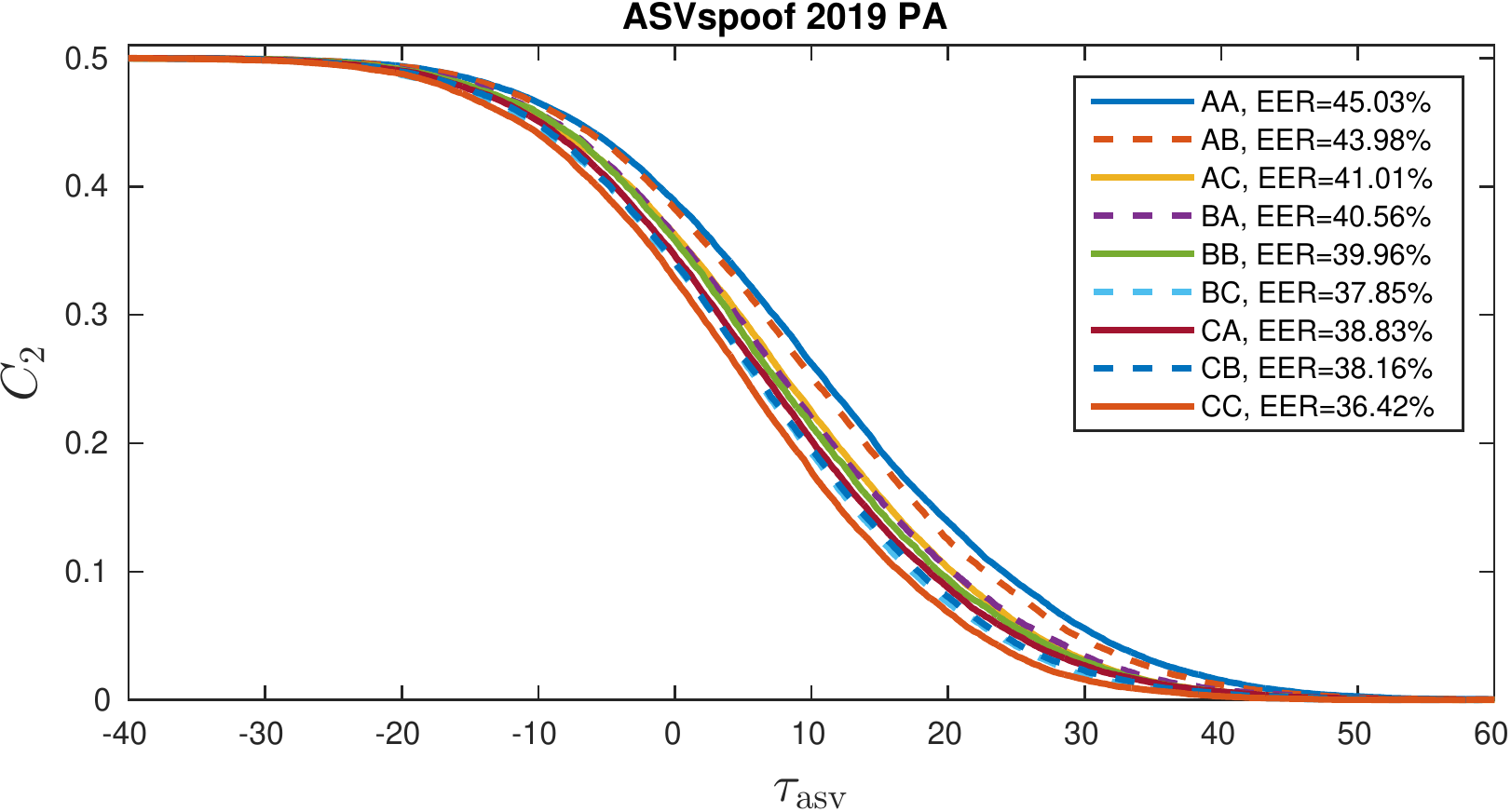}
    \caption{$C_2$ over the ASV threshold $\tau_{\text{asv}}$ per spoofing attack type, on the ASVspoof 2019 logical access (upper panel) and physical access (lower panel) scenarios. The indicated EERs are those of the x-vector based ASV system.}
    \label{fig:empirical-C2-ASVspoof19}
\end{figure}

Thus far, we have focused on scores pooled from all attacks, even if their effectiveness (in terms of fooling ASV) varies. It is therefore useful to diagnose attack-specific, empirical $C_2$ functions, similar to those in Fig.~\ref{fig:asv-cond-coeffs} (note that $C_0$ and $C_1$ depend on target and nontarget trials only). The attack-specific $C_2$ graphs are shown in Fig.~\ref{fig:empirical-C2-ASVspoof19} along with the corresponding ASV EERs. We observe, first, that different attacks produce similarly-shaped but differently located graphs along the $\tau_\text{asv}$ axis. For fixed $\tau_\text{asv}$, the more effective attack (\emph{i.e.}, higher EER) gives larger $C_2$. Second, there is substantial variation of $C_2$, especially on the LA condition. For the PA condition, both the $C_2$ graphs and the EERs vary less across attacks. Unlike the LA attacks generated by a large number of researchers and teams, the PA attacks were generated through a common simulation model with a few control parameters only~\cite{wang2019asvspoof}, which may explain the more homogenous behavior. The interested reader may refer to \cite{wang2019asvspoof,Todisco2019} for further details on ASVspoof 2019 attack generation and their impact upon ASV.

\begin{table}[]
\caption{ASV-constrained t-DCF obtained using different ASV and CM thresholds. The first two lines indicate the empirical criterion (and data) to set $\tau_\text{asv}$ and $\tau_\text{cm}$, respectively.} 

\centering
\setlength\tabcolsep{1.5pt}
\begin{tabular}{|>{\color{black}}m{1.3cm}||>{\color{black}}P{1.7cm}|>{\color{black}}P{1.7cm}|>{\color{black}}P{1.7cm}||>{\color{black}}P{1.7cm}|}
\hline
Fix $\tau_\text{asv} \rightarrow$ & EER (eval)                & EER (dev)                  & $\min C_0$ (dev)               & $\min C_0$ (dev)     \\
\hline
Fix $\tau_\text{cm} \rightarrow$  &   $\min \text{t-DCF}$ (eval) &  $\min \text{t-DCF}$ (eval) &  $\min \text{t-DCF}$ (eval) &  $\min \text{t-DCF}$ (dev) \\
\hline\hline
System              & \multicolumn{3}{>{\color{black}}c||}{$\min$ t-DCF}                                                   & actual t-DCF      \\
\hline
\multicolumn{5}{|>{\color{black}}c|}{\textbf{ASVspoof 2019 LA}}                                                                             \\
\hline
ASV floor           & 0.0627                   & 0.0860                    & 0.0304                    & 0.0304          \\
\hline
T05                 & 0.0692                   & 0.0923                    & 0.0367                    & 0.3260          \\
T45                 & 0.1104                   & 0.1327                    & 0.0785                    & 0.4989           \\
\hline
\multicolumn{5}{|>{\color{black}}c|}{\textbf{ASVspoof 2019 PA}}                                                                                     \\
\hline
ASV floor          & 0.1354                   & 0.1389                    & 0.0628                    & 0.0628          \\
\hline
T28                 & 0.1437                   & 0.1472                    & 0.0715                    & 0.0718          \\
T45                 & 0.1460                   & 0.1495                    & 0.0740                    & 0.0769 \\
\hline        
\end{tabular}
\label{tab:operation-points-ASVspoof19}
\end{table}

\subsection{Empirical Threshold Selection Using t-DCF}\label{sec:minActual}

Until this point, we considered an arbitrary $\tau_\text{asv}$ (set at the EER operating point) along with an optimum $\tau_\text{cm}$. These were the choices in the ASVspoof 2019 challenge though the former is not aligned with the t-DCF specification. Further, we have considered oracle calibration only where both thresholds are set on the evaluation data. Thus, in our final experiment we demonstrate use of the t-DCF to guide selection of both thresholds (on development and evaluation data).  

In line with the ASV-constrained approach, we consider a particular scenario where the ASV and CM systems developers agree upon a specific t-DCF parametrization (specified by the bank) but optimise their respective systems separately, while sharing ASV error rates only. Using common development data, they proceed as follows:
    \begin{enumerate}
        \item Since the ASV system is not designed to reject spoofing attacks and hence by using target/non-target trials only, the ASV system developer optimizes $\tau_\text{asv}$ by minimizing the ASV floor, i.e.:
            \begin{equation}\label{eq:select-asv-threshold}
                \begin{aligned}
                \tau_\text{asv}^* & = \arg\min_{\tau_\text{asv}} C_0(\tau_\text{asv})\\
                    & = \arg\min_{\tau_\text{asv}} \Big\{ \pi_\text{tar}C_\text{miss}P_\text{miss}^\text{asv}(\tau_\text{asv})+\pi_\text{non}C_\text{fa}P_\text{fa}^\text{asv}(\tau_\text{asv})\Big\}.
                \end{aligned}
            \end{equation}
        \item Using the shared ASV error rates, bonafide (target/non-target) and spoofed trials, the CM developer determines $C_0$, $C_1$, and $C_2$ coefficients in \eqref{eq:tdcf-coefficients} and sets the CM to operate at the empirical minimum of the ASV-constrained t-DCF in \eqref{eq:tDCF-ASV-constrained}:
            \begin{equation}
                \tau_\text{cm}^* = \arg\min_{\tau_\text{cm}} \text{t-DCF}(\tau_\text{cm}).
            \end{equation}
    \end{enumerate}

We contrast the above approach with the EER-based ASV threshold selection (as used in previous experiments) in Table~\ref{tab:operation-points-ASVspoof19} which shows results for the evaluation partitions of the ASVspoof 2019 datasets for LA and PA tasks. The first two columns corresponds to the EER-based optimization of $\tau_\text{asv}$ on evaluation and development data, respectively. The last two columns correspond to choosing the ASV threshold using \eqref{eq:select-asv-threshold} on development data. 

Concerning $\tau_\text{cm}$, in turn, the first three columns correspond to oracle calibration of the CM (minimum t-DCF on the evaluation set). The last column corresponds to the \textbf{actual} t-DCF where both $\tau_\text{asv}$ and $\tau_\text{cm}$ are set on the development set. Results are shown for the top-2 systems for the ASVspoof 2019 LA and PA scenarios. Similar to Fig.~\ref{fig:tDCF_allDBs}, the lower bound (ASV floor) is also shown for reference purposes.

Upon comparison of results in the first two columns we see that, for the LA scenario, the ASV EER threshold set on development data is suboptimal compared to the ASV EER threshold tuned on evaluation data; there are differences between results in each column. In contrast, for the PA scenario, results are similar, no matter if the threshold is tuned on the development data or the evaluation data. Results in column 3 indicate that the ASV threshold set to minimize $C_0$ on the development data leads to substantially lower t-DCF values than in columns 1 and 2. This is expected since the EER  operating point represents a miscalibrated ASV system that is generally not intended as a minimizer of the t-DCF. 

Finally, results for the LA scenario in column 4 indicate that a CM threshold tuned on the development set does not generalize well to the evaluation set; except for the perfect CM (ASV floor), results in columns~4 are substantially worse than those in column~3.
In contrast, the difference is comparatively small for the PA scenario. This might be due to the same reasons noted in Fig.~\ref{fig:empirical-C2-ASVspoof19} --- the diversity in the spoofing attacks for the LA scenario is greater than that for the PA scenario.  Data for the latter was generated with a common simulation procedure, leading to more homogeneuos attacks.

These results demonstrate the potential of t-DCF as an empirical threshold selection criterion. The above procedure is intended as a demonstration that follows the format of the ASVspoof 2019 challenge, though there are a number of open questions that we discuss below.

\section{Discussion and Future Work}

Before concluding, we discuss here the assumed  independence in the t-DCF metric, and outline a number of open research problems exposed in this work.

\subsection{The ASV---CM Independence Assumption}\label{subsec:asv-cm-independence}

The formulation of the tandem error rates in Eqs.~\eqref{eq:tandem-error-probabilities} is based upon the assumption that ASV and CM system error rates are independent. While this may seem somewhat questionable, there are a number of reasons that support such a restrictive assumption. They relate to the specific ASVspoof scenario and the fundamental differences between 
ASV and CM systems:
\begin{enumerate}
    \item as discussed in Section \ref{sec:the-two-systems}, ASV and CM systems address different detection tasks and hence the two detectors provide complementary views, even to the same data;
    \item in a speaker-independent CM setting, the two systems are trained using disjoint speakers; 
    \item ASV systems provide scores for each (target speaker, test) pair whereas speaker-independent CM systems, in turn, use one anti-spoofing model to score all test utterances;
    \item the ASV and the CM systems could be developed by different researchers, teams or companies and can hence be based on different ideas, methods, software implementations, control parameters, and training data choices;
    \item the two systems typically use different features (e.g. CQCC vs. MFCC features) and classifier back-ends.
\end{enumerate}
Thus, \emph{before} application of the tandem system to evaluation data, ASV and CM scores can be treated as being independent, as can the respective nontarget/target and bonafide/spoof score distributions. \emph{After} the same tandem system is applied to evaluation data, however, ASV and CM cannot necessarily be treated as being independent --- there will be some \emph{conditional} dependence by virtue of both systems being executed on the same data. We nonetheless assert in \eqref{eq:tandem-error-probabilities} that ASV and CM scores are conditionally \emph{in}dependent. 

The primary reason for our conditional independence assumption is practical: it allows the ASV and CM error rates be computed by different parties (and from different data). We have deemed this as a necessity given the practical focus of the ASVspoof challenge series. Nonetheless, if the joint distribution of ASV and CM scores \emph{is} available (for instance, when the same person develops both systems), it may be useful to assess the impact of statistical dependency on the estimated detection error rates. Independence assumptions are sometimes difficult to avoid, e.g.~the well known NIST SREs~\cite{Greenberg2020_twenty_years} which assume statistically independent trials, yet reuse many times data from the same target/non-target speakers~\cite{Wu2017-impact-of-data-dependency}. A deeper study of the in/dependence issue is kept for future work.

\subsection{The Need for a Tandem Calibration Model}

Aside from experiments reported in Section~\ref{sec:minActual}, the issue of optimally calibrated detectors is largely overlooked in this paper.  In doing so, we have sidestepped the important but very real problem of threshold setting. The default practice in ASV research, with widespread acceptance by the ASV research community, is to fix $\tau_\text{asv}$ to the Bayes minimum-risk operating point~\cite{DudaHartStork01}, with the assumption that detection scores are well-calibrated log-likelihood ratios (LLRs)~\cite{LeeuwenB07}. The Bayes threshold is given analytically by the DCF parameters as $\tau_\text{asv}=\log [(C_\text{fa}/C_\text{miss})\cdot (1-\pi_\text{tar})/\pi_\text{tar}]$. Since arbitrary detectors may yield poorly calibrated LLRs 
it is customary to apply a \emph{calibration transform} in the score domain~\cite{Leeuwen2013-calibration,Cumani2019_tied,Cumani2019NormalVM}. Among other benefits, this allows calibrated scores to be used readilly with different DCF parameters. Furthermore, calibration allows principled decomposition of \emph{discrimination loss} (how bad the system is in terms of classification; at an ideal threshold) and \emph{calibration loss} (how badly off the threshold is from that ideal threshold) \cite{LeeuwenB07,Brummer-application-independent}.
The apparent benefits of calibration may cause the reader to wonder why we did not apply a tandem calibration model in the case of the t-DCF. 

The main reason for having avoided calibration is that the manner by which it should be applied in the tandem setting is far from being a simple extension of calibration in the case of the DCF.  Under the tandem framework, we have not only one additional \emph{system} (CM) but also one additional \emph{class} (spoof). At this point, the authors have no analytical expressions for Bayes-optimal ASV and CM thresholds.  In addition, the unconstrained and ASV-constrained cases may require different treatments. The ASV-constrained case yields a cost function \emph{for particular spoofing attacks} whose impact upon the ASV system is known; in reality, we do not know that impact in advance. In summary, how calibration transforms should be defined for tandem systems remains an open question and one that deserves attention in future work.

\section{Conclusions}

The intention of the authors has been to provide a self-contained tutorial on the tandem detection cost (t-DCF) framework that generalizes the standard DCF. Extending upon \cite{Kinnunen2018-tDCF}, our special focus has been on a \emph{constrained} t-DCF formulation, where the biometric system (here, ASV) is essentially treated as a black-box. The constrained cost serves as a guide for the optimization of a countermeasure for a given biometric system. A variant of the constrained t-DCF was put to its first stress test in the latest ASVspoof 2019 challenge~\cite{wang2019asvspoof} and we anticipate it remaining as the primary metric in future editions of the challenge.

The question of how the performance of \emph{any} binary classifier is to be assessed is much more subtle than it appears on the surface; it took some considerable time for the DCF framework to be absorbed as an integral part of ASV system development --- and even longer to migrate from \emph{ad-hoc} `threshold optimization' recipes to calibrated log-likelihood ratios \cite{Leeuwen2013-calibration,Brummer-application-independent,Brummer2010-PhD,Cumani2019NormalVM,Cumani2019_tied}. We hope that the current study serves to reduce the risk of similarly slow adoption of application-directed metrics within the anti-spoofing community. Since the necessity for tandem systems (consisting of two subsystems) add to the complexity of the assessment issue, the authors have purposefully left out a number of related topics, such as calibration; we focused on \emph{minimum} t-DCF (with oracle threshold). We plan to address calibration in our future work.

The authors note that a variety of different, \emph{adhoc} metrics remain popular in the assessment of biometric systems (beyond the voice trait). Presentation attack detection is a relatively recent, but growing and evolving area of research, and this state of rapid development may go some way to explain the lack of application-directed metrics in use today. What is clear, however, is that next-generation biometric systems \emph{must} be prepared for the possibility of spoofing (whether it be ever-improving DeepFake video and synthetic speech quality, or potential fraud in high-stakes applications including border control and forensics). To help prepare for a future where biometrics may no longer be trusted, we need meaningful metrics both for performace assessment and optimization. One benefit of the proposed t-DCF framework lies in its \emph{generality}. While, on account of the authors' research interests, voice biometrics has been the running example, the t-DCF itself requires nothing beyond the detection scores (or hard decisions) of the biometric recognizer and the presentation attack detector. The specification of cost parameters is left to the domain expert.

\section*{Acknowledgements}

This work has been sponsored by Academy of Finland (proj. no. 309629), Japan Science and Technology (JST), and the Department of Defense under Air Force Contract FA8721-05-C-0002. The work has also been partially funded by the ANR-DFG French-German RESPECT project and the JST-ANR VoicePersonae project.  Opinions, interpretations, conclusions and recommendations are those of the authors and are not necessarily endorsed by the United States Government.

\section*{Appendix: Gaussian score simulator}\label{appendix:numerical-sim}

In \cite{Leeuwen2013-calibration}, relations between Gaussian score distributions, well-calibrated log-likelihood ratios (LLRs) and the EER were drawn to derive a supervised score calibration recipe under the assumption of Gaussian nontarget/target scores. In this special case, the means of non/target distributions are symmetric and also relate to their variance. Such constrained score distributions can be parameterized by a single number, such as the EER. In the following, we outline a simple score simulator models for target, nontarget, and spoof classes. 

Our score simulator models all the class-conditional score distributions in \eqref{eq:asv-probability-densities} and \eqref{eq:cm-probability-densities} as Gaussians,
    \begin{equation}\label{eq:general-score-sim}
        \begin{aligned}
            p_R(r|\theta_\bullet)& =\mathcal{N}(r|\mu^\text{asv}_\bullet,(\sigma_\bullet^2)^\text{asv}),  & \bullet & \in \{\text{tar},\text{non},\text{spoof}\}\\
            p_Q(q|\theta_\circ)& =\mathcal{N}(q|\mu^\text{cm}_\circ,(\sigma_\circ^2)^\text{cm}), & \circ & \in \{\text{bona},\text{spoof}\},
        \end{aligned}
    \end{equation}
where $\mathcal{N}(\cdot|\mu,\sigma^2)$ denotes the univariate normal density with mean $\mu$ and variance $\sigma^2$. For the CM, the target and nontarget distributions are assumed to be the same --- the CM cannot discriminate between target and nontarget classes.

As there are three classes within the ASV system and two classes within the CM system, \eqref{eq:general-score-sim} requires the specification of $(3 + 2) \times 2 = 10$ parameters. A convenient means to reduce the number is through the EERs of each system --- target vs. nontarget EER for ASV and bonafide vs. spoof EER for CM.
The EER can be expressed analytically in terms of Gaussian parameters~\cite{Poh2004-multistream,Leeuwen2013-calibration}. Even if the tail behavior of empirical ASV scores obtained from typical back-ends~\cite{prince2007probabilistic} differs from that of Gaussians~\cite{Cumani2019NormalVM,Cumani2019_tied}, supervised calibration using constrained Gaussians~\cite{Leeuwen2013-calibration} leads to near-optimum calibration near the minimum cost operating point~\cite{Cumani2019NormalVM,Cumani2019_tied}.

Let us begin with the ASV system with $\mathcal{N}(r|\mu_\text{tar}^\text{asv},(\sigma_\text{tar}^2)^\text{asv})$ and $\mathcal{N}(r|\mu_\text{non}^\text{asv},(\sigma_\text{non}^2)^\text{asv})$ as the target and nontarget score distributions, respectively. The analytic EER is given by $P_\text{e}=1-\Phi(F^\text{asv})$, where $\Phi(\cdot)$ is the cumulative distribution function of the standard normal distribution and  $F^\text{asv}=(\mu_\text{tar}^\text{asv}-\mu_\text{non}^\text{asv})/(\sigma_\text{tar}^\text{asv} + \sigma_\text{non}^\text{asv})$. For completely overlapped distributions with equal means, one obtains the chance level $P_\text{e}^\text{asv} =\frac{1}{2}$. The four parameters collapse to a scalar $F^\text{asv}$, which uniquely specifies the EER (as $\Phi$ is bijective). We \emph{tie} the means and variances so that we have only one degree of freedom, specified by the EER, from which we determine the four parameters. To this end, we adopt the approach of \cite{Leeuwen2013-calibration} with shared variance $\sigma^\text{asv} \equiv\sigma_\text{tar}^\text{asv}=\sigma_\text{non}^\text{asv}$, symmetric means ($\mu^\text{asv} \equiv\mu_\text{tar}^\text{asv}=-\mu_\text{non}^\text{asv}$) and mean and variance being related by $\sigma^\text{asv} = \sqrt{2\mu^\text{asv}}$. To sum up, $p_R(r|\theta_\text{tar})=\mathcal{N}(r|\mu^\text{asv},2\mu^\text{asv})$,  $p_R(r|\theta_\text{non})=\mathcal{N}(r|-\mu^\text{asv},2\mu^\text{asv})$.

\begin{figure}[!t]
	\centering
  \includegraphics[width=0.80\columnwidth]{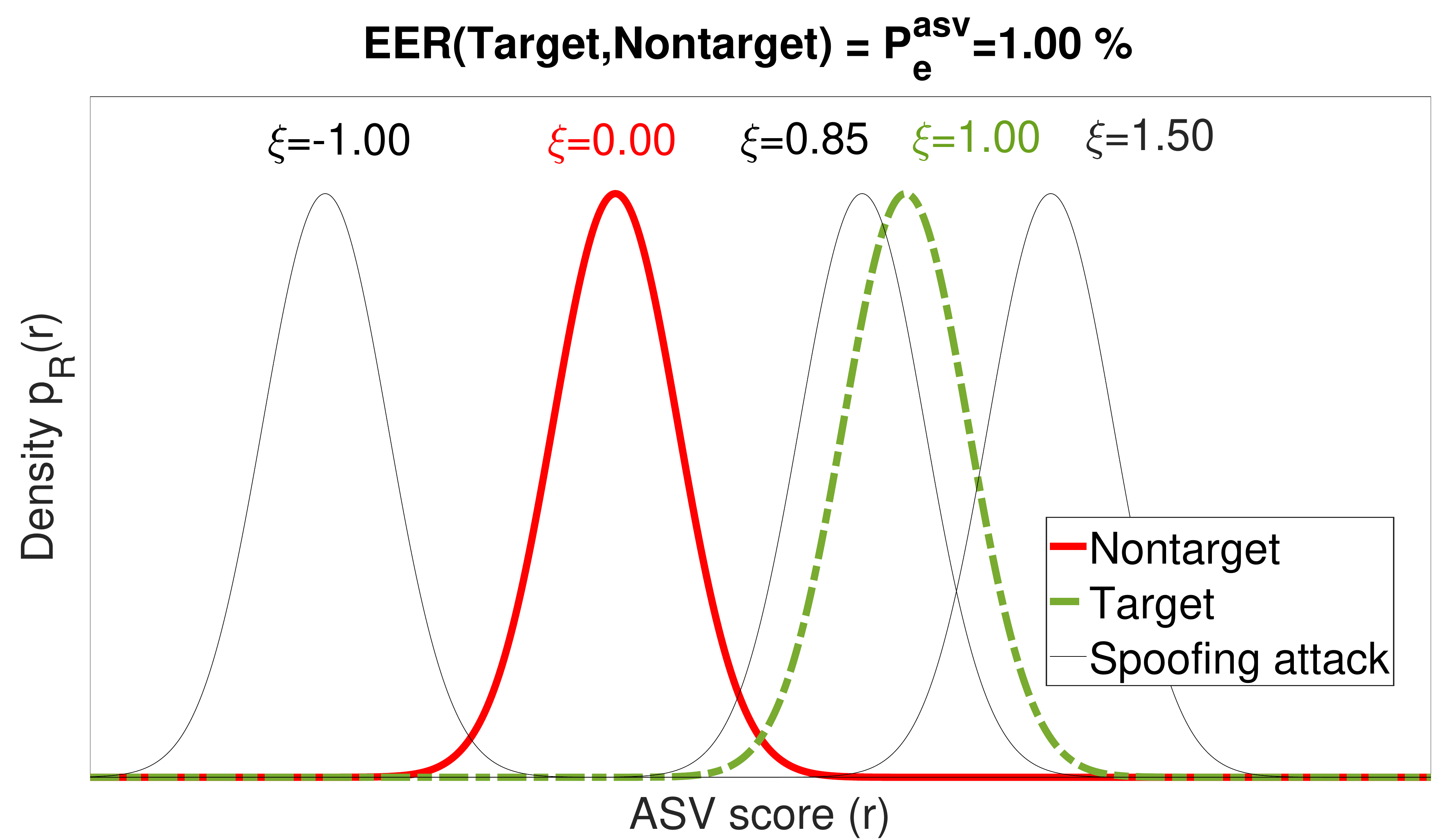}
	\caption{Simulated, Gaussian-distributed ASV score distributions with varied spoofing factor, $\xi$. Nontarget and target classes are special cases with $\xi=0$ and $\xi=1$, respectively.}
	\label{fig:spoofing-factor}
\end{figure}

After having determined the target and nontarget distributions from a given EER, $P_\text{e}^\text{asv}$, we proceed by defining the ASV spoof score distribution as:
    \begin{equation}\label{eq:appendix-pR_spoof}
        p_R(r|\theta_\text{spoof}) = \mathcal{N}\Big(r|\mu^\text{asv}\left(2\xi - 1\right), 2\mu^\text{asv}\Big),
    \end{equation}
characterized by an additional parameter $\xi \in \mathbb{R}$ that we dub as the \emph{spoofing factor}. It is illustrated in Fig.~\ref{fig:spoofing-factor}, with the following interpretation:
    \begin{itemize}
        \item $\xi=1$ implies spoof mean equal to target mean (attack indistinguishable from the target speaker).
        \item $\xi=0$ implies spoof mean equal to nontarget mean, \emph{i.e.} zero-effort spoofing attack.
        \item $\xi > 1$ implies spoof mean higher than target mean.
        \item $\xi < 0$ implies spoof mean less than nontarget mean.
    \end{itemize}
The  typical case is $0 \ll \xi < 1$, \emph{i.e.} attacks that produce substantially higher scores than nontargets but do not quite reach the target scores due to modeling imperfections, difficulties in gathering spoofing attack training data, or other reasons.

The CM bona fide score distributions are specified in the same way as for the ASV system: given a desired bonafide-to-spoof EER, $P_\text{e}^\text{cm}$, we find the distributions of bona fide and spoof classes as $\mathcal{N}(\mu^\text{cm},2\mu^\text{cm})$ and 
$\mathcal{N}(-\mu^\text{cm},2\mu^\text{cm})$, respectively, following the same parameter constraints noted above.

To state our simulator assumptions in an alternative way, the ASV and CM score random variables are assumed to be statistically independent (for practical reasons; see Section \ref{subsec:asv-cm-independence}). 
The joint probability density functions of target, nontarget and spoof classes are then the product of their marginal distributions, which can be represented as bi-variate Gaussians with a diagonal covariance matrix.
For the target distribution, for instance, we have
    \begin{equation}
        \begin{aligned}
        p_{R,Q}(r,q|\theta_\text{tar}) & = \mathcal{N}(r|\mu^\text{asv}_\text{tar},(\sigma_\text{tar}^2)^\text{asv})\,\mathcal{N}(q|\mu^\text{cm}_\text{bona},(\sigma_\text{bona}^2)^\text{cm})\\
        & =\mathcal{N}(\vec{\mu}_\text{tar},\mtx{\Sigma}_\text{tar})\nonumber,
        \end{aligned}
    \end{equation}   
with 
    \begin{equation}
        \vec{\mu}_\text{tar} = 
        \begin{bmatrix}
        \mu^\text{asv}\\
        \mu^\text{cm}\\
        \end{bmatrix},
        \;\;\;\;\;
        \mtx{\Sigma}_\text{tar} = 
        \begin{bmatrix}
        2\mu^\text{asv} & 0\\
        0 & 2\mu^\text{cm}\\
        \end{bmatrix}.\nonumber
    \end{equation}
Similarly, for the nontarget and spoof class we have
    \begin{equation}
        \vec{\mu}_\text{non} = 
        \begin{bmatrix}
        -\mu^\text{asv}\\
        \mu^\text{cm}\\
        \end{bmatrix},
        \;\;\;\;\;
        \mtx{\Sigma}_\text{non} = 
        \begin{bmatrix}
        2\mu^\text{asv} & 0\\
        0 & 2\mu^\text{cm}\\
        \end{bmatrix}\nonumber
    \end{equation}
    \begin{equation}
        \vec{\mu}_\text{spoof} = 
        \begin{bmatrix}
        \mu^\text{asv}(2\xi -1)\\
        -\mu^\text{cm}\\
        \end{bmatrix},
        \;\;\;\;\;
        \mtx{\Sigma}_\text{spoof} = 
        \begin{bmatrix}
        2\mu^\text{asv} & 0\\
        0 & 2\mu^\text{cm}\\
        \end{bmatrix}.\nonumber
    \end{equation}

In summary, the three control parameters of our score simulator that define the above distributions, are:
    \begin{enumerate}
        \item Target-to-nontarget EER of ASV, $P_\text{e}^\text{asv}$, as a model of the discrimination performance of ASV;
        \item ASV spoofing factor, $\xi \in \mathbb{R}$, as a model of how effective the spoofing attacks are in fooling the ASV;
        \item Bonafide-to-spoof EER of CM, $P_\text{e}^\text{cm}$, as a model of CM discrimination performance. 
    \end{enumerate}

Under the above Gaussian model, the detection error rates in \eqref{eq:asv-cm-detection-errors} are given by,
\begin{equation}
    \begin{aligned}
    P_\text{miss}^\text{asv}(\tau_\text{asv})& =\Phi\left(\frac{\tau_\text{asv}-\mu^\text{asv}}{\sqrt{2\mu^\text{asv}}}\right)\\
    P_\text{fa}^\text{asv}(\tau_\text{asv})&=1-\Phi\left(\frac{\tau_\text{asv}+\mu^\text{asv}}{\sqrt{2\mu^\text{asv}}}\right)\\
    P_\text{fa,spoof}^\text{asv}(\tau_\text{asv})&=1-\Phi\left(\frac{\tau_\text{asv}-\mu^\text{asv}\left(2\xi - 1\right)}{\sqrt{2\mu^\text{asv}}}\right)\\
    P_\text{miss}^\text{cm}(\tau_\text{cm})& =\Phi\left(\frac{\tau_\text{cm}-\mu^\text{cm}}{\sqrt{2\mu^\text{cm}}}\right)\\
    P_\text{fa}^\text{cm}(\tau_\text{cm})&=1-\Phi\left(\frac{\tau_\text{cm}+\mu^\text{cm}}{\sqrt{2\mu^\text{cm}}}\right),\\
    \end{aligned}
\end{equation}
where $\Phi(\tau)$ denotes the cumulative distribution function (CDF) of the standard normal distribution. The standardization operator $\tau \mapsto (\tau - \mu)/\sigma$ yields the CDF of a nonstandard normal distribution with mean $\mu$ and variance $\sigma^2$.

\bibliographystyle{IEEEtran}
\bibliography{tDCF_BibEntries}

\begin{thebibliography}{10}
\providecommand{\url}[1]{#1}
\csname url@samestyle\endcsname
\providecommand{\newblock}{\relax}
\providecommand{\bibinfo}[2]{#2}
\providecommand{\BIBentrySTDinterwordspacing}{\spaceskip=0pt\relax}
\providecommand{\BIBentryALTinterwordstretchfactor}{4}
\providecommand{\BIBentryALTinterwordspacing}{\spaceskip=\fontdimen2\font plus
\BIBentryALTinterwordstretchfactor\fontdimen3\font minus
  \fontdimen4\font\relax}
\providecommand{\BIBforeignlanguage}[2]{{%
\expandafter\ifx\csname l@#1\endcsname\relax
\typeout{** WARNING: IEEEtran.bst: No hyphenation pattern has been}%
\typeout{** loaded for the language `#1'. Using the pattern for}%
\typeout{** the default language instead.}%
\else
\language=\csname l@#1\endcsname
\fi
#2}}
\providecommand{\BIBdecl}{\relax}
\BIBdecl

\bibitem{Greenberg2020_twenty_years}
\BIBentryALTinterwordspacing
C.~S. Greenberg, L.~P. Mason, S.~O. Sadjadi, and D.~A. Reynolds, ``Two decades
  of speaker recognition evaluation at the {National Institute of Standards and
  Technology},'' \emph{Computer Speech {\&} Language}, vol.~60, 2020. [Online].
  Available: \url{https://doi.org/10.1016/j.csl.2019.101032}
\BIBentrySTDinterwordspacing

\bibitem{Doddington2000-NIST-overview}
\BIBentryALTinterwordspacing
G.~R. Doddington, M.~A. Przybocki, A.~F. Martin, and D.~A. Reynolds, ``The
  {NIST} speaker recognition evaluation --- {Overview}, methodology, systems,
  results, perspective,'' \emph{Speech Communication}, vol.~31, no. 2-3, pp.
  225--254, 2000. [Online]. Available:
  \url{https://doi.org/10.1016/S0167-6393(99)00080-1}
\BIBentrySTDinterwordspacing

\bibitem{DudaHartStork01}
R.~O. Duda, P.~E. Hart, and D.~G. Stork, \emph{Pattern Classification},
  2nd~ed.\hskip 1em plus 0.5em minus 0.4em\relax New York: Wiley, 2001.

\bibitem{Jaynes03}
E.~T. Jaynes, \emph{Probability theory: The logic of science}.\hskip 1em plus
  0.5em minus 0.4em\relax Cambridge: Cambridge University Press, 2003.

\bibitem{isopad}
{ISO/IEC 30107-1:2016}, ``{Information technology — Biometric presentation
  attack detection — Part 1: Framework},''
  \url{https://www.iso.org/obp/ui/#iso:std:iso-iec:30107:-1:ed-1:v1:en}, 2016,
  [Online; accessed 22-February-2018].

\bibitem{Interspeech2013SS}
N.~{E}vans, T.~{K}innunen, and J.~{Y}amagishi, ``{S}poofing and countermeasures
  for automatic speaker verification,'' in \emph{Proc. Interspeech}, 2013, pp.
  925--929.

\bibitem{Satoh2001-HMM-imposture}
\BIBentryALTinterwordspacing
T.~Satoh, T.~Masuko, T.~Kobayashi, and K.~Tokuda, ``A robust speaker
  verification system against imposture using an {HMM}-based speech synthesis
  system,'' in \emph{Proc. {EUROSPEECH}}, 2001, pp. 759--762. [Online].
  Available:
  \url{http://www.isca-speech.org/archive/eurospeech\_2001/e01\_0759.html}
\BIBentrySTDinterwordspacing

\bibitem{Sizov2015-tifs}
\BIBentryALTinterwordspacing
A.~Sizov, E.~Khoury, T.~Kinnunen, Z.~Wu, and S.~Marcel, ``Joint speaker
  verification and antispoofing in the i-vector space,'' \emph{{IEEE} Trans.
  Information Forensics and Security}, vol.~10, no.~4, pp. 821--832, 2015.
  [Online]. Available: \url{https://doi.org/10.1109/TIFS.2015.2407362}
\BIBentrySTDinterwordspacing

\bibitem{Sahid2016-integrated}
\BIBentryALTinterwordspacing
M.~Sahidullah, H.~Delgado, M.~Todisco, H.~Yu, T.~Kinnunen, N.~Evans, and
  Z.~Tan, ``Integrated spoofing countermeasures and automatic speaker
  verification: An evaluation on {ASVspoof} 2015,'' in \emph{Proc.
  Interspeech}, 2016, pp. 1700--1704. [Online]. Available:
  \url{https://doi.org/10.21437/Interspeech.2016-1280}
\BIBentrySTDinterwordspacing

\bibitem{Todisco2018-integrated}
\BIBentryALTinterwordspacing
M.~Todisco, H.~Delgado, K.~A. Lee, M.~Sahidullah, N.~Evans, T.~Kinnunen, and
  J.~Yamagishi, ``Integrated presentation attack detection and automatic
  speaker verification: Common features and {Gaussian} back-end fusion,'' in
  \emph{Proc. Interspeech}, 2018, pp. 77--81. [Online]. Available:
  \url{http://dx.doi.org/10.21437/Interspeech.2018-2289}
\BIBentrySTDinterwordspacing

\bibitem{Kinnunen2018-tDCF}
\BIBentryALTinterwordspacing
T.~Kinnunen, K.~A. Lee, H.~Delgado, N.~Evans, M.~Todisco, M.~Sahidullah,
  J.~Yamagishi, and D.~A. Reynolds, ``{t-DCF}: a detection cost function for
  the tandem assessment of spoofing countermeasures and automatic speaker
  verification,'' in \emph{Proc. Odyssey}, 2018, pp. 312--319. [Online].
  Available: \url{http://dx.doi.org/10.21437/Odyssey.2018-44}
\BIBentrySTDinterwordspacing

\bibitem{Bishop2006}
C.~M. Bishop, \emph{Pattern Recognition and Machine Learning (Information
  Science and Statistics)}.\hskip 1em plus 0.5em minus 0.4em\relax Berlin,
  Heidelberg: Springer-Verlag, 2006.

\bibitem{LeeuwenB07}
\BIBentryALTinterwordspacing
D.~A. van Leeuwen and N.~Br{\"{u}}mmer, ``An introduction to
  application-independent evaluation of speaker recognition systems,'' in
  \emph{Speaker Classification {I:} Fundamentals, Features, and Methods}, 2007,
  pp. 330--353. [Online]. Available:
  \url{https://doi.org/10.1007/978-3-540-74200-5\_19}
\BIBentrySTDinterwordspacing

\bibitem{Brummer2010-PhD}
N.~Br\"ummer, ``Measuring, refining and calibrating speaker and language
  information extracted from speech,'' Ph.D. dissertation, Stellenbosch
  University, 2010.

\bibitem{Wu2015-asvspoof}
\BIBentryALTinterwordspacing
Z.~Wu, T.~Kinnunen, N.~Evans, J.~Yamagishi, C.~Hanil{\c{c}}i, M.~Sahidullah,
  and A.~Sizov, ``{ASVspoof} 2015: the first automatic speaker verification
  spoofing and countermeasures challenge,'' in \emph{Proc. Interspeech}, 2015,
  pp. 2037--2041. [Online]. Available:
  \url{http://www.isca-speech.org/archive/interspeech_2015/i15_2037.html}
\BIBentrySTDinterwordspacing

\bibitem{Kinnunen2017assessing}
T.~Kinnunen, M.~Sahidullah, H.~Delgado, M.~Todisco, N.~Evans, J.~Yamagishi, and
  K.~A. Lee, ``The {ASVspoof} 2017 challenge: Assessing the limits of replay
  spoofing attack detection,'' in \emph{Proc. Interspeech}, 2017, pp. 2--6.

\bibitem{Todisco2019}
M.~Todisco, X.~Wang, V.~Vestman, M.~Sahidullah, H.~Delgado, A.~Nautsch,
  J.~Yamagishi, N.~Evans, T.~H. Kinnunen, and K.~A. Lee, ``{ASVspoof 2019:
  Future Horizons in Spoofed and Fake Audio Detection},'' in \emph{Proc.
  Interspeech}, 2019, pp. 1008--1012.

\bibitem{Snyder2018XVectorsRD}
D.~Snyder, D.~Garcia-Romero, G.~Sell, D.~Povey, and S.~Khudanpur, ``X-vectors:
  Robust {DNN} embeddings for speaker recognition,'' in \emph{Proc. ICASSP},
  2018, pp. 5329--5333.

\bibitem{prince2007probabilistic}
S.~J. Prince and J.~H. Elder, ``Probabilistic linear discriminant analysis for
  inferences about identity,'' in \emph{2007 IEEE 11th International Conference
  on Computer Vision}.\hskip 1em plus 0.5em minus 0.4em\relax IEEE, 2007, pp.
  1--8.

\bibitem{povey2011kaldi}
D.~Povey, A.~Ghoshal, G.~Boulianne, L.~Burget, O.~Glembek, N.~Goel,
  M.~Hannemann, P.~Motlicek, Y.~Qian, P.~Schwarz \emph{et~al.}, ``The {Kaldi}
  speech recognition toolkit,'' IEEE Signal Processing Society, Tech. Rep.,
  2011.

\bibitem{nagrani2017voxceleb}
\BIBentryALTinterwordspacing
A.~Nagrani, J.~S. Chung, and A.~Zisserman, ``{VoxCeleb:} a large-scale speaker
  identification dataset,'' in \emph{Proc. Interspeech}, 2017, pp. 2616--2620.
  [Online]. Available: \url{http://dx.doi.org/10.21437/Interspeech.2017-950}
\BIBentrySTDinterwordspacing

\bibitem{wang2019asvspoof}
\BIBentryALTinterwordspacing
X.~Wang, J.~Yamagishi, M.~Todisco, H.~Delgado, A.~Nautsch, N.~Evans,
  M.~Sahidullah, V.~Vestman, T.~Kinnunen, K.~A. Lee, L.~Juvela, P.~Alku, Y.-H.
  Peng, H.-T. Hwang, Y.~Tsao, H.-M. Wang, S.~L. Maguer, M.~Becker,
  F.~Henderson, R.~Clark, Y.~Zhang, Q.~Wang, Y.~Jia, K.~Onuma, K.~Mushika,
  T.~Kaneda, Y.~Jiang, L.-J. Liu, Y.-C. Wu, W.-C. Huang, T.~Toda, K.~Tanaka,
  H.~Kameoka, I.~Steiner, D.~Matrouf, J.-F. Bonastre, A.~Govender, S.~Ronanki,
  J.-X. Zhang, and Z.-H. Ling, ``{ASVspoof} 2019: {A} large-scale public
  database of synthesized, converted and replayed speech,'' \emph{Computer
  Speech \& Language}, vol.~64, p. 101114, 2020. [Online]. Available:
  \url{http://www.sciencedirect.com/science/article/pii/S0885230820300474}
\BIBentrySTDinterwordspacing

\bibitem{Wu2017-impact-of-data-dependency}
J.~C. {Wu}, A.~F. {Martin}, C.~S. {Greenberg}, and R.~N. {Kacker}, ``The impact
  of data dependence on speaker recognition evaluation,'' \emph{IEEE/ACM
  Transactions on Audio, Speech, and Language Processing}, vol.~25, no.~1, pp.
  5--18, 2017.

\bibitem{Leeuwen2013-calibration}
\BIBentryALTinterwordspacing
D.~A. van Leeuwen and N.~Br{\"{u}}mmer, ``The distribution of calibrated
  likelihood-ratios in speaker recognition,'' in \emph{Proc. Interspeech},
  2013, pp. 1619--1623. [Online]. Available:
  \url{http://www.isca-speech.org/archive/interspeech_2013/i13_1619.html}
\BIBentrySTDinterwordspacing

\bibitem{Cumani2019_tied}
S.~{Cumani} and P.~{Laface}, ``Tied normal variance–mean mixtures for linear
  score calibration,'' in \emph{Proc. ICASSP}, May 2019, pp. 6121--6125.

\bibitem{Cumani2019NormalVM}
S.~Cumani, ``Normal variance-mean mixtures for unsupervised score
  calibration,'' in \emph{Proc. Interspeech}, 2019, pp. 401--405.

\bibitem{Brummer-application-independent}
N.~Br\"ummer and J.~du~Preez, ``Application-independent evaluation of speaker
  detection,'' \emph{Computer Speech \& Language}, vol.~20, no.~2, pp.
  230--275, 2006.

\bibitem{Poh2004-multistream}
\BIBentryALTinterwordspacing
N.~Poh and S.~Bengio, ``Why do multi-stream, multi-band and multi-modal
  approaches work on biometric user authentication tasks?'' in \emph{Proc.
  ICASSP}, 2004, pp. 893--896. [Online]. Available:
  \url{https://doi.org/10.1109/ICASSP.2004.1327255}
\BIBentrySTDinterwordspacing

\end{thebibliography}

%

\begin{IEEEbiography}[{\includegraphics[width=1in,height=1.25in,clip,keepaspectratio]{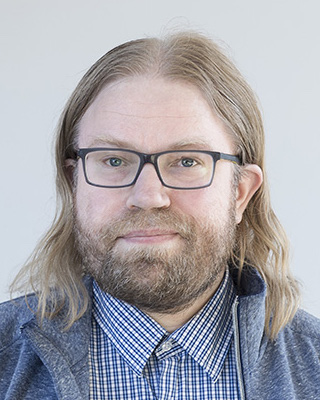}}]{Tomi H. Kinnunen} is an Associate Professor at the University of Eastern Finland. He received his Ph.D. degree in computer science from the University of Joensuu in 2005. From 2005 to 2007, he was an Associate Scientist at the Institute for Infocomm Research (I2R), Singapore. Since 2007, he has been with UEF. From 2010-2012, he was funded by a postdoctoral grant from the Academy of Finland. He has been a PI or co-PI in three other large Academy of Finland-funded projects and a partner in the H2020-funded OCTAVE project. He chaired the \emph{Odyssey} workshop in 2014. From 2015 to 2018, he served as an Associate Editor for IEEE/ACM Trans. on Audio, Speech and Language Processing and from 2016 to 2018 as a Subject Editor in \emph{Speech Communication}. In 2015 and 2016, he visited the National Institute of Informatics, Japan, for 6 months under a mobility grant from the Academy of Finland, with a focus on voice conversion and spoofing. Since 2017, he has been Associate Professor at UEF, where he leads the Computational Speech Group. He is one of the cofounders of the ASVspoof challenge, a nonprofit initiative that seeks to evaluate and improve the security of voice biometric solutions under spoofing attacks.
\end{IEEEbiography}
%
\begin{IEEEbiography}[{\includegraphics[width=1in,height=1.25in,clip,keepaspectratio]{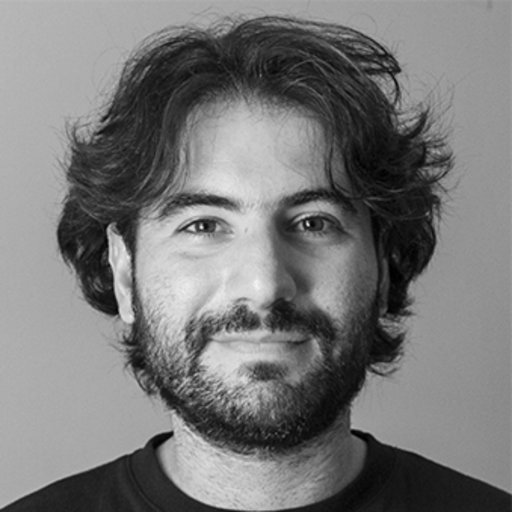}}]{H\'ector Delgado} received his Ph.D. degree in Telecommunication and System Engineering from the Autonomous University of Barcelona (UAB), Spain, in 2015. From 2015 to 2019 he was with the Speech and Audio Processing Research Group at EURECOM (France). Since 2019 he is a Senior Research Scientist at Nuance Communications Inc. He serves as an associate editor for the EURASIP Journal on Audio, Speech, and Music Processing. He is a co-organiser of the ASVspoof challenge since its 2017 edition. His research interests include signal processing and machine learning applied to speaker recognition and
diarization, speaker recognition anti-spoofing and audio segmentation.
\end{IEEEbiography}
%
\begin{IEEEbiography}[{\includegraphics[width=1in,height=1.25in,clip,keepaspectratio]{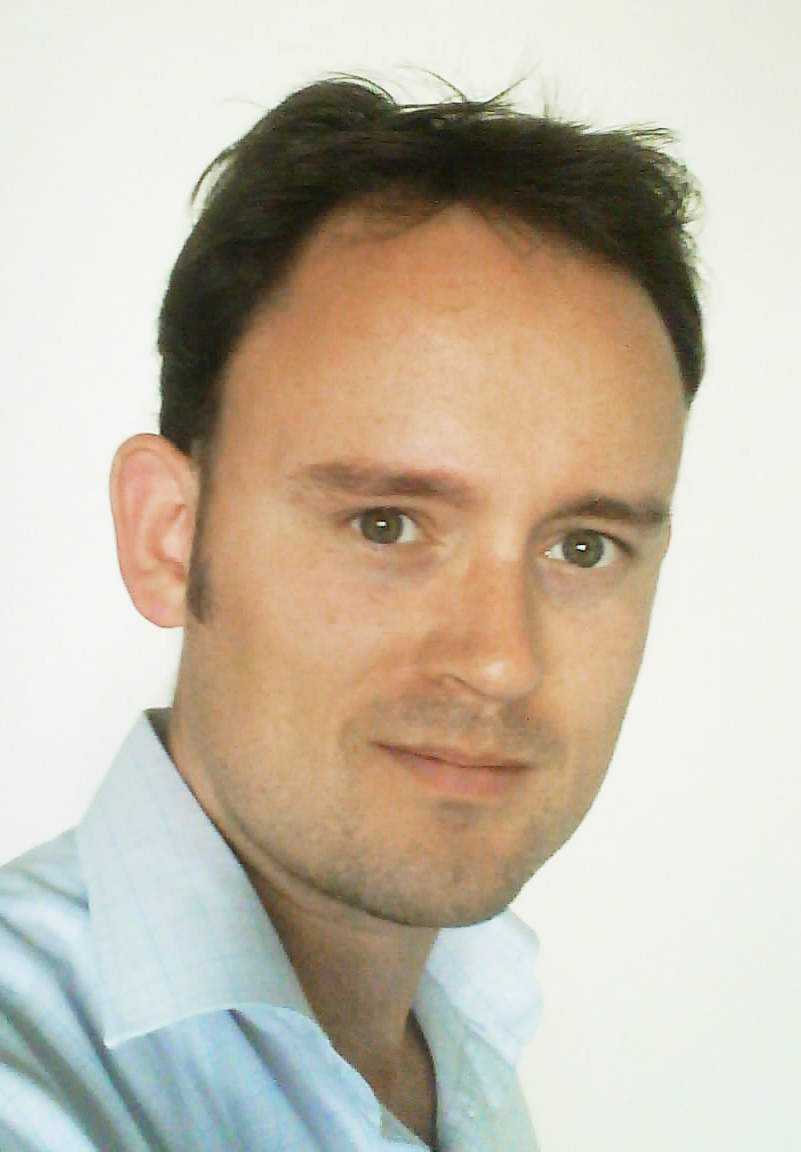}}]{Nicholas Evans} is a Professor at EURECOM, France, where he heads research in Audio Security and Privacy. He is a co-founder of the community-led, ASVspoof Challenge series and has lead or co-lead a number of special issues and sessions with an anti-spooing theme. He participated in the EU FP7 Tabula Rasa and H2020 OCTAVE projects, both involving anti-spoofing. Today, his team is leading the EU H2020 TReSPAsS-ETN project, a training initiative in security and privacy for multiple biometric traits.  He co-edited the second edition of the Handbook of Biometric Anti-Spoofing, served previously on the IEEE Speech and Language Technical Committee and serves currently as an asscociate editor for the IEEE Trans.\ on Biometrics, Behavior, and Identity Science.
\end{IEEEbiography}
%
\begin{IEEEbiography}
[{\includegraphics[width=1.2in,height=1.0in,clip,keepaspectratio]{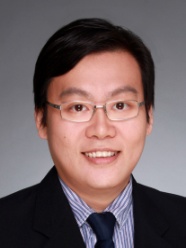}}]
{Kong Aik Lee} (M'05-SM'16) is currently a Senior Principal Researcher at the Biometrics Research Laboratories, NEC Corp., Japan. He received his Ph.D. degree from Nanyang Technological University, Singapore, in 2006. From 2006 to 2018, he was a Scientist at the Human Language Technology department, I$^2$R, A*STAR, Singapore, where he led the speaker recognition group. He was the recipient of Singapore IES Prestigious Engineering Achievement Award 2013 for his contribution to voice biometrics technology. He serves as an Editorial Board Member for Elsevier Computer Speech and Language (2016 - present), and an Associate Editor for IEEE/ACM Transactions on Audio, Speech and Language Processing (2017 - present). He is an elected member of IEEE Speech and Language Technical Committee. He chairs thte Speaker Odyssey 2020 Workshop.
\end{IEEEbiography}
\begin{IEEEbiography}[{\includegraphics[width=1in,height=1.25in,clip,keepaspectratio]{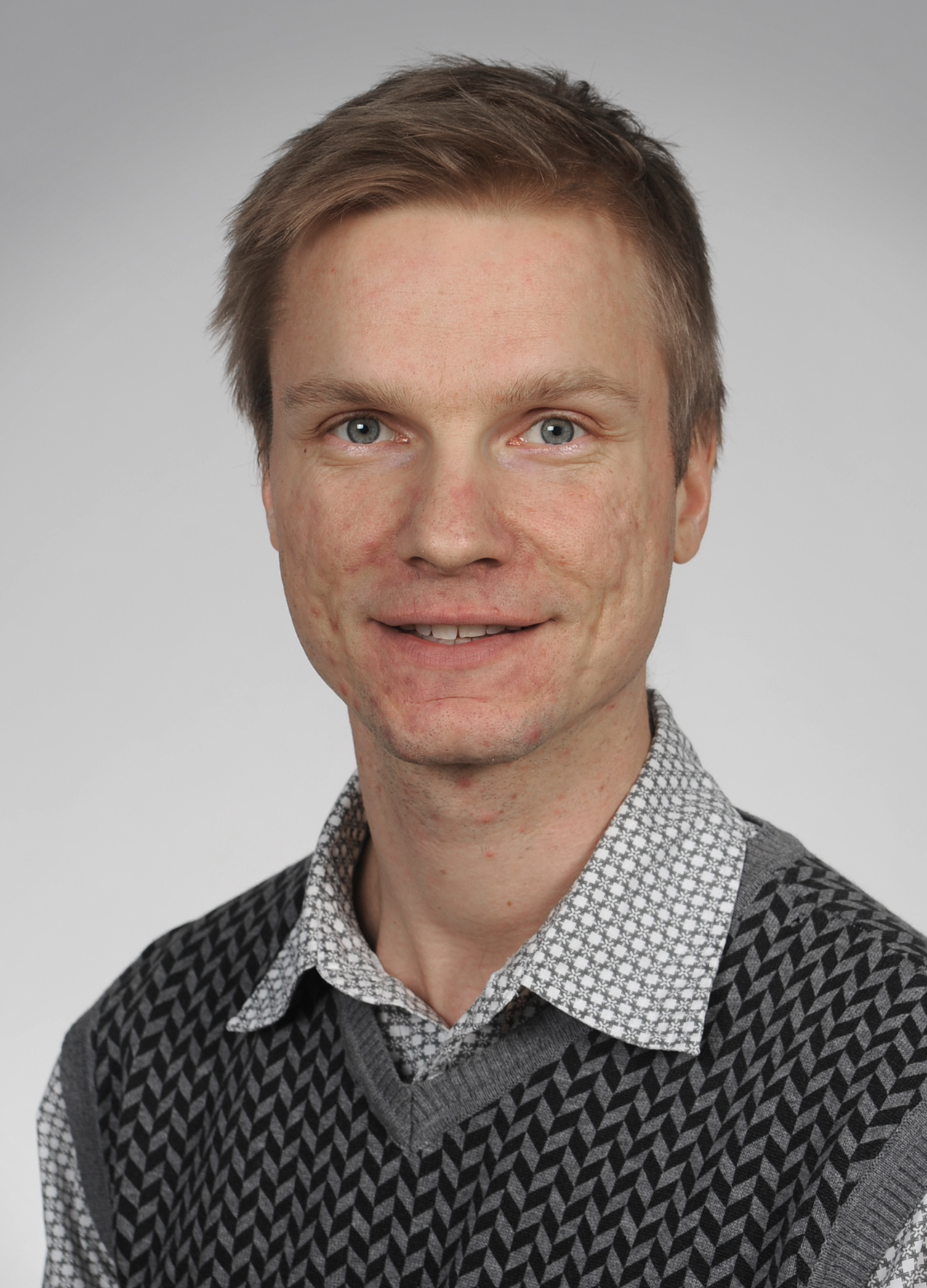}}]{Ville Vestman} is an Early Stage Researcher at the University of Eastern Finland (UEF). He received his M.S. degree in mathematics from UEF in 2013. Since 2015, his research work at UEF has been focused on speech technology and, more specifically, on speaker recognition. He is one of the co-organizers of the ASVspoof 2019 challenge.
\end{IEEEbiography}

\begin{IEEEbiography}[{\includegraphics[width=1in,height=1.25in,clip,keepaspectratio]{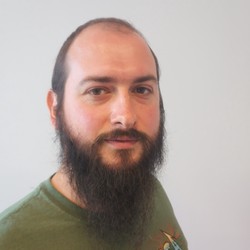}}]{Andreas Nautsch} is with the Audio Security and Privacy research group (EURECOM). He received the doctorate from Technische Universit\"{a}t Darmstadt in 2019, where he was with the biometrics group within the German National Research Center for Applied Cybersecurity. He received B.Sc. and M.Sc. degrees from Hochschule Darmstadt (dual studies with atip GmbH) in 2012 and 2014, respectively. He served as an expert delegate to ISO/IEC and as project editor of the ISO/IEC 19794-13:2018 standard. Andreas is a co-initiator and secretary of the ISCA Special Interest Group on Security \& Privacy in Speech Communication.
\end{IEEEbiography}

\begin{IEEEbiography}[{\includegraphics[width=1in,height=1.25in,clip,keepaspectratio]{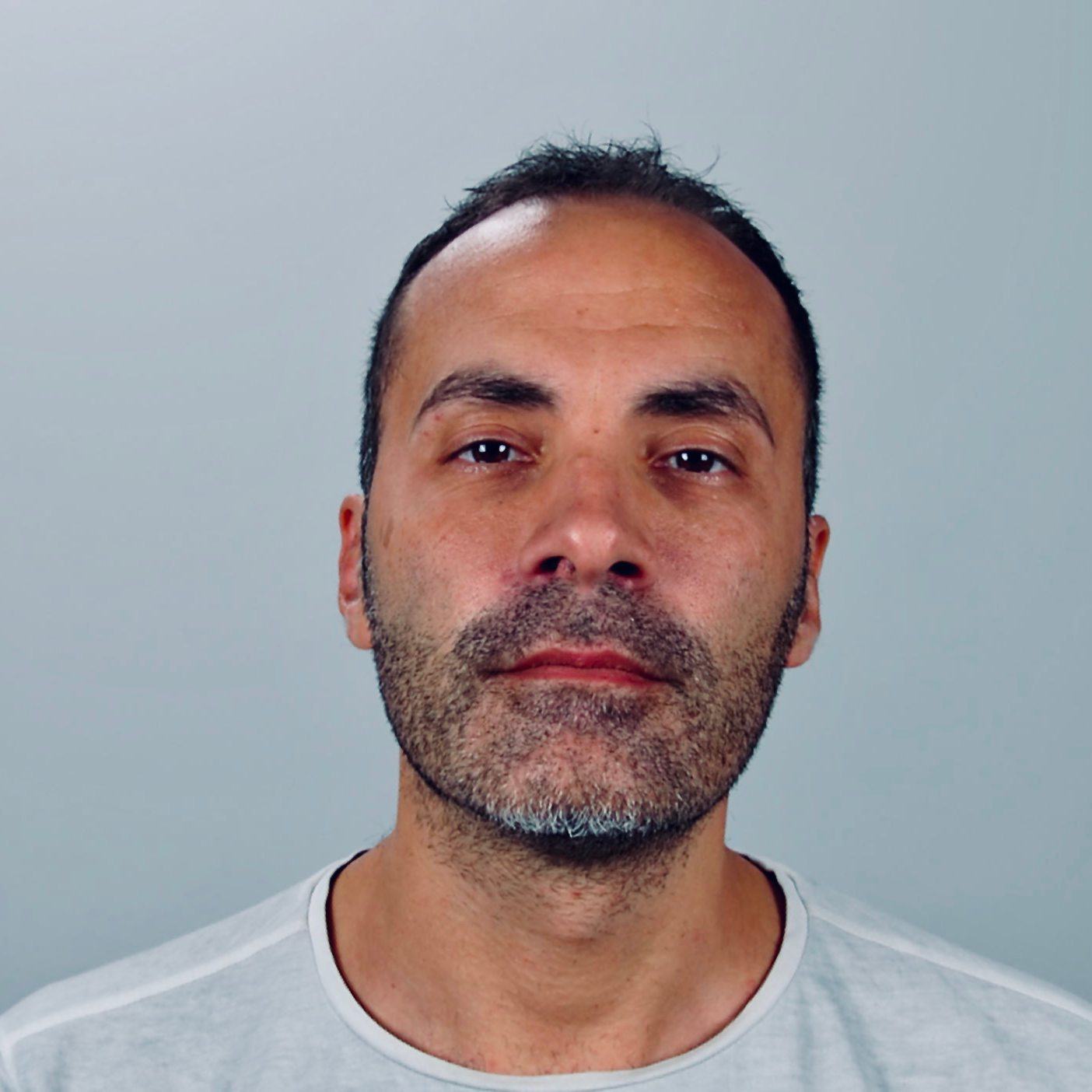}}]{Massimiliano Todisco} is an Assistant Professor within the Digital Security Department at EURECOM, France. He received his Ph.D. degree in Sensorial and Learning Systems Engineering from the University of Rome Tor Vergata in 2012. Currently, he is serving as principal investigator and coordinator for TReSPAsS-ETN, a H2020 Marie Skłodowska-Curie Innovative Training Network (ITN) and RESPECT, a PRCI project funded by the French ANR and the German DFG. He co-organises the ASVspoof challenge series, which is community-led challenges which promote the development of countermeasures to protect automatic speaker verification (ASV) from the threat of spoofing. He is the inventor of constant Q cepstral coefficients (CQCC), the most commonly used anti-spoofing features for speaker verification and first author of the highest-cited technical contribution in the field in the last three years. He has more than 90 publications. His current interests are in developing end-to-end architectures for speech processing and speaker recognition, fake audio detection and anti-spoofing, and the development of privacy preservation algorithms for speech signals based on encryption solutions that support computation upon signals, templates and models in the encrypted domain.\end{IEEEbiography}
%
%
\begin{IEEEbiography}[{\includegraphics[width=1in,height=1.25in,clip,keepaspectratio]{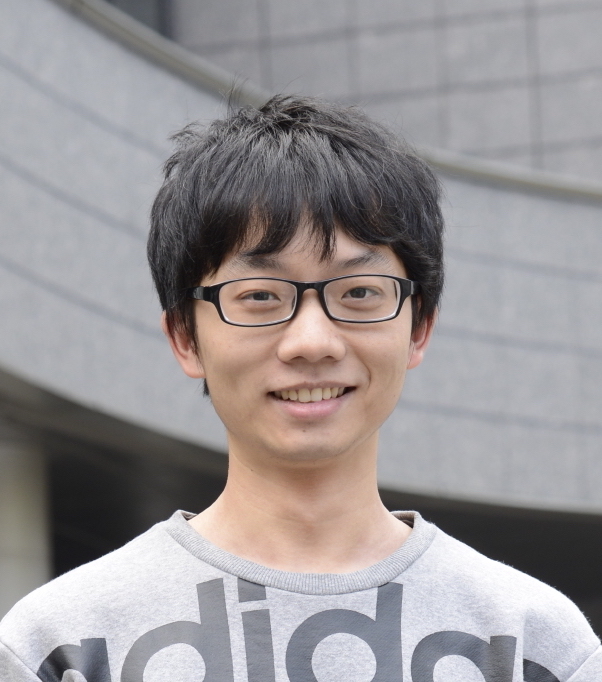}}]{Xin Wang} (S'16 - M'18)
is a project researcher at National Institute of Informatics, Japan. He received the Ph.D. degree from SOKENDAI, Japan, in 2018. Before that, he received M.S. and B.E degrees from University of Science and Technology of China and University of Electronic Science and Technology of China in 2015 and 2012, respectively. His research interests include statistical speech synthesis and machine learning.
\end{IEEEbiography}
%
%
\begin{IEEEbiography}[{\includegraphics[width=1in,height=1.25in,clip,keepaspectratio]{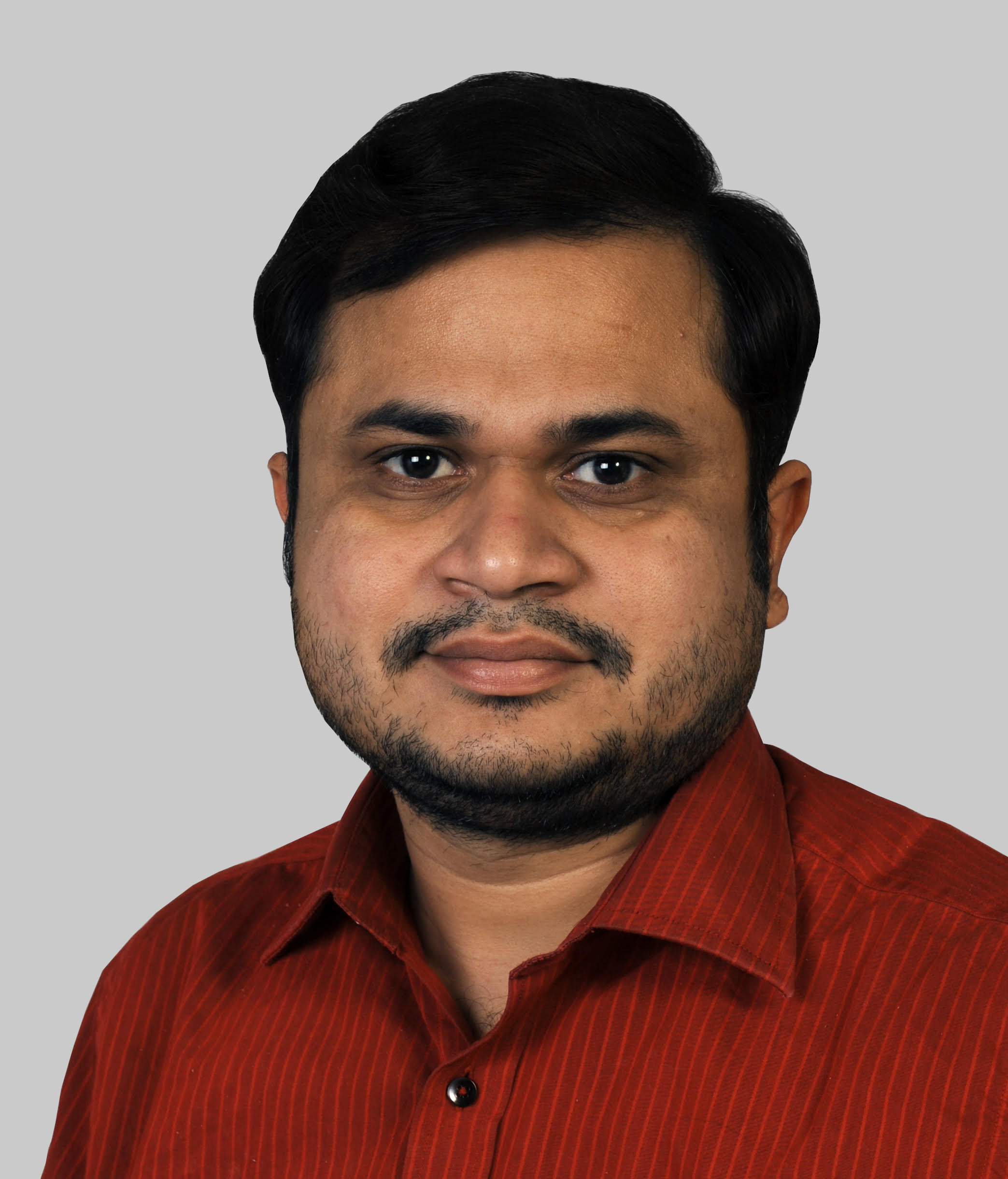}}]{Md Sahidullah} (S’09, M'15) received his Ph.D. degree in the area of speech processing from the Department of Electronics \& Electrical Communication Engineering, Indian Institute of Technology Kharagpur in 2015. Prior to that he obtained the Bachelors of Engineering degree in Electronics and Communication Engineering from Vidyasagar University in 2004 and the Masters of Engineering degree in Computer Science and Engineering from West Bengal University of Technology in 2006. In 2014-2017, he was a postdoctoral researcher with the School of Computing, University of Eastern Finland. In January 2018, he joined MULTISPEECH team, Inria, France as a post-doctoral researcher where he currently holds a starting research position. His research interest includes robust speaker recognition and spoofing countermeasures. He is also part of the organizing team of two Automatic Speaker Verification Spoofing and Countermeasures Challenges: ASVspoof 2017 and ASVspoof 2019. Presently, he is also serving as Associate Editor for the IET Signal Processing and Circuits, Systems, and Signal Processing.
\end{IEEEbiography}
%
%
\begin{IEEEbiography}[{\includegraphics[width=1in,height=1.25in,clip,keepaspectratio]{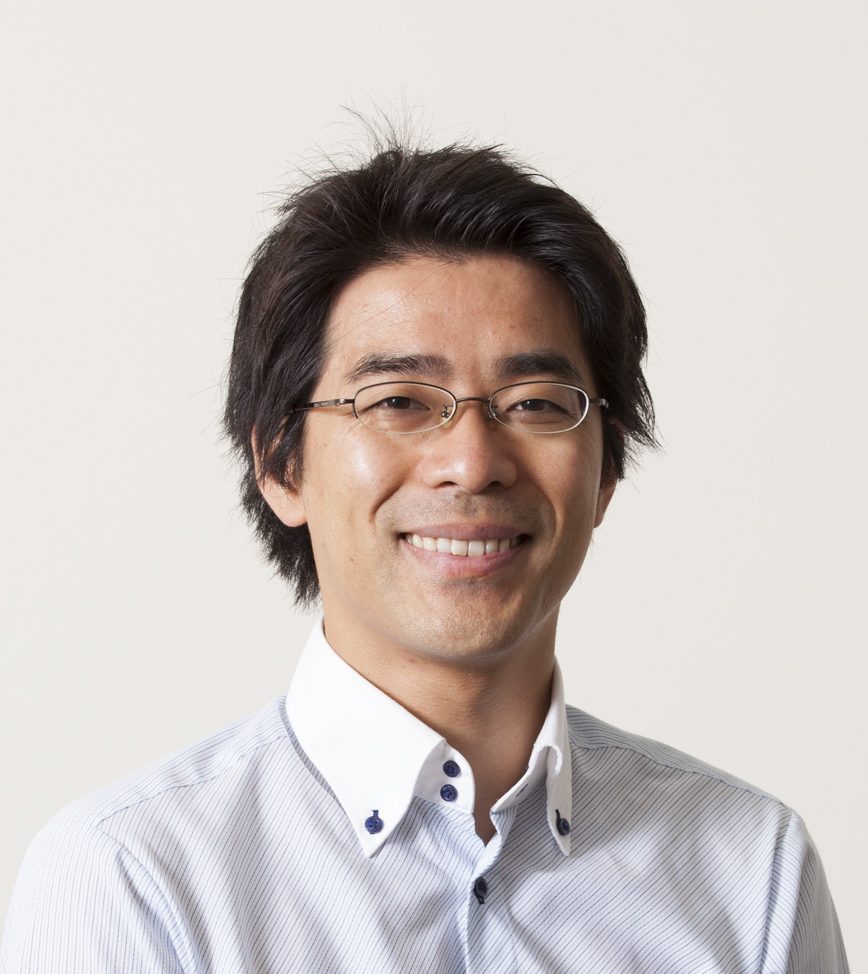}}]{Junichi Yamagishi} (SM'13) is a professor at National Institute of Informatics in Japan. He is also a senior research fellow in the Centre for Speech Technology Research (CSTR) at the University of Edinburgh, UK. He was awarded a Ph.D.\ by Tokyo Institute of Technology in 2006 for a thesis that pioneered speaker-adaptive speech synthesis and was awarded the Tejima Prize as the best Ph.D.\ thesis of Tokyo Institute of Technology in 2007. Since 2006, he has authored and co-authored over 250 refereed papers in international journals and conferences. He was awarded the Itakura Prize from the Acoustic Society of Japan, the Kiyasu Special Industrial Achievement Award from the Information Processing Society of Japan, and the Young Scientists’ Prize from the Minister of Education, Science and Technology, the JSPS prize, the Docomo mobile science award in 2010, 2013, 2014, 2016, and 2018, respectively. He served previously as co-organizer for the bi-annual ASVspoof special sessions at INTERSPEECH 2013-9, the bi-annual Voice conversion challenge at INTERSPEECH 2016 and Odyssey 2018, an organizing committee member for the 10th ISCA Speech Synthesis Workshop 2019 and a technical program committee member for IEEE ASRU 2019. He also served as a member of the IEEE Speech and Language Technical Committee, as an Associate Editor of the IEEE/ACM TASLP and a Lead Guest Editor for the IEEE JSTSP SI on Spoofing and Countermeasures for Automatic Speaker Verification. He is currently a PI of JST-CREST and ANR supported VoicePersonae project. He also serves as a chairperson of ISCA SynSIG and as a Senior Area Editor of the IEEE/ACM TASLP.
\end{IEEEbiography}
%
\begin{IEEEbiography}[{\includegraphics[width=1in,height=1.25in,clip,keepaspectratio]{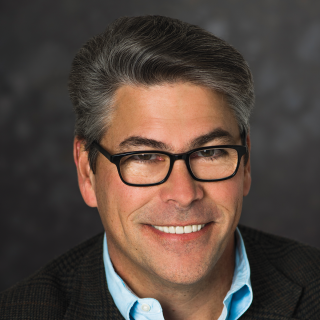}}]{Douglas A. Reynolds} (S'85–M'86–SM'98–F'10) is a senior member of the technical staff at MIT Lincoln Laboratory where he provides technical oversight of the projects in speaker and language recognition and speech-content-based information retrieval. Dr. Reynolds received his PhD from the Georgia Institute of Technology in 1992 with a dissertation on applying Gaussian Mixture Models (GMMs) to automatic speaker recognition. His current research is focused on application of speech technology to real-world scenarios and domain adaptation of speech systems. Dr. Reynolds is a Fellow of the IEEE, recipient of the 2017 MIT Lincoln Laboratory Technical Excellence Award, and a founding member of the Odyssey Speaker Recognition Workshop series.
\end{IEEEbiography}
\vfill





\end{document}